\documentclass[usenatbib]{mnras}\usepackage{graphics,graphicx,epsfig,ulem,multirow,appendix,lscape,amsmath,amssymb,color,enumitem,natbib,url}

\usepackage{upgreek}

\usepackage{bm} 

\hypersetup{draft}

\mathchardef\shorthyphen="2D                                            
\newcommand{\mEarth}{\; {\rm M_\oplus}}                         
\newcommand{\mEarthPerYr}{\; {\rm M_\oplus \; yr^{-1}}}                    
\newcommand{\au}{\; {\rm au}}                                           
\newcommand{\pc}{\; {\rm pc}}                                           
\newcommand{\rSun}{\; {\rm R_\odot}}                            
\newcommand{\m}{\; {\rm m}}                                                     
\newcommand{\cm}{\; {\rm cm}}                                                     
\newcommand{\mm}{\; {\rm mm}}                                                     
\newcommand{\nm}{\; {\rm nm}}                                                     
\newcommand{\km}{\; {\rm km}}                                           
\newcommand{\um}{\; {\rm \mu m}}                                        
\newcommand{\mas}{\; {\rm mas}}                                           
\newcommand{\yr}{\; {\rm yr}}                                         
\newcommand{\myr}{\; {\rm Myr}}                                         
\newcommand{\gyr}{\; {\rm Gyr}}                                         
\newcommand{\jy}{\; {\rm Jy}}                                                     
\newcommand{\K}{\; {\rm K}}                                                     
\newcommand{\gPerCmCubed}{\; {\rm g \; cm^{-3}}}        
\newcommand{\percent}{\; {\rm per \; cent}}                     
\newcommand{\perYr}{\; \rm yr^{-1}}           
\newcommand{\cometsPerYr}{\; \rm comets \; yr^{-1}}           
\newcommand{\HD}{{\rm HD} \;}                                           

\newcommand{\change}[1]{{#1}}

\long\def\symbolfootnote[#1]#2{\begingroup%
\def\thefootnote{\fnsymbol{footnote}}\footnote[#1]{#2}\endgroup} 
\def\apjl{ApJ}
\def\apjs{ApJS}
\def\aap{A\&A}

\begin{document}

\title[Cometary supply of hot dust]
  {Hot exozodis: cometary supply without trapping is unlikely to be the mechanism}
\author[T. D. Pearce et al.]
  {Tim D. Pearce$^1$\thanks{timothy.pearce@uni-jena.de},
  Florian Kirchschlager$^2$,
  Ga\change{\"{e}}l Rouill\'{e}$^3$,
  Steve Ertel$^{4, 5}$,
  Alexander Bensberg$^6$,\newauthor
  Alexander V. Krivov$^1$,
  Mark Booth$^1$,
  Sebastian Wolf$^6$\change{,}
    \change{Jean-Charles Augereau$^7$}\\
  $^1$Astrophysikalisches Institut und Universit\"{a}tssternwarte, Friedrich-Schiller-Universit\"{a}t Jena, Schillerg\"{a}{\ss}chen 2-3, D-07745 Jena,\\ Germany\\
  $^2$Sterrenkundig Observatorium, Ghent University, Krijgslaan 281 - S9, 9000 Gent, Belgium\\
  $^3$Laboratory Astrophysics Group of the Max Planck Institute for Astronomy at the Friedrich-Schiller-Universit\"{a}t Jena, Institute of\\Solid State Physics, Helmholtzweg 3, D-07743 Jena, Germany\\
  $^4$Large Binocular Telescope Observatory, \change{The University of Arizona,} 933 North Cherry Ave, Tucson, AZ 85721, USA\\
  $^5$Department of Astronomy and Steward Observatory, \change{The University of Arizona,} 933 North Cherry Ave, Tucson, AZ 85721, USA\\
 $^6$Institute of Theoretical Physics and Astrophysics, University of Kiel, Leibnizstra{\ss}e 15, 24118 Kiel, Germany\\
 \change{$^7$Univ. Grenoble Alpes, CNRS, IPAG, 38000, Grenoble, France}\\
	  }
\date{Released 2002 Xxxxx XX}
\pagerange{\pageref{firstpage}--\pageref{lastpage}} \pubyear{2002}

\def\LaTeX{L\kern-.36em\raise.3ex\hbox{a}\kern-.15em
    T\kern-.1667em\lower.7ex\hbox{E}\kern-.125emX}

\newtheorem{theorem}{Theorem}[section]

\label{firstpage}

\maketitle


\begin{abstract}
Excess near-infrared emission is detected around one fifth of main-sequence stars, but its nature is a mystery. These excesses are interpreted as \change{thermal emission from} populations of small, hot dust very close to their stars (`hot exozodis'), but such grains should rapidly sublimate or be blown out of the system. To date, no model has fully explained this phenomenon. One mechanism commonly suggested in the literature is cometary supply, where star-grazing comets deposit dust close to the star, replenishing losses from grain sublimation and blowout. However, we show that this mechanism alone is very unlikely to be responsible for hot exozodis. We model the trajectory and size evolution of dust \change{grains} released by star-grazing comets, to establish the dust and \change{comet} properties required to reproduce hot-exozodi observations. We find that cometary supply alone can only reproduce observations if dust ejecta has an extremely steep size distribution upon release, and the dust-deposition rate is extraordinarily high. These requirements strongly contradict our current understanding of cometary dust and planetary systems. Cometary supply is therefore \change{unlikely} to be solely responsible for hot exozodis, so \change{may} need to be combined with some dust-trapping mechanism (such as gas or magnetic trapping) if it is to reproduce observations.
\end{abstract}


\section{Introduction}
\label{sec: introduction}

Excess near-infrared (NIR) emission is detected at the ${\sim1\percent}$ level around one \change{f}ifth of main sequence stars, across a diverse range of stellar types and ages \citep{Absil2006, Absil2013, Ertel2014, Mennesson2014, Ertel2016, Nunez2017, Absil2021}. These excesses are attributed to hot dust in close proximity to stars, so are often referred to as `hot exozodis'. Interferometric and polarization measurements suggest this dust primarily comprises sub-micron grains with a steep size distribution, at distances comparable to those where carbon and silicates are expected to sublimate \citep{diFolco2007, Akeson2009, Defrere2011, Defrere2012, Lebreton2013, Marshall2016, Kirchschlager2017}. It appears that grains are carbonaceous rather than silicate-rich \citep{Absil2006, Kirchschlager2017, Sezestre2019}, and there are no clear correlations between the presence of NIR excesses and excesses at mid-\change{infrared} (MIR) or far-\change{infrared} (FIR) \change{wavelengths} \citep{MillanGabet2011, Ertel2014, Mennesson2014, Ertel2018, Ertel2020, Absil2021}. In many cases, strong NIR emission is detected whilst no MIR emission is found.

The nature of hot dust is a mystery, because such small, hot grains should rapidly sublimate or blow away from stars. It is very unlikely that grains are replenished \textit{in situ} via steady-state collisional cascades \citep{Wyatt2007, Lebreton2013}, so a series of more exotic scenarios have been proposed to explain hot\change{-}dust populations. These hypothesise that hot dust is continually resupplied from elsewhere in the system and/or trapped near the star. However, no model has been able to fully explain hot exozodis and their ubiquity across different star types and ages.

An early attempt at an explanation was the Poynting-Robertson (PR)-drag pileup model, where grains migrate inwards from some distant dust source until they approach the star and sublimate, then blow out of the system \citep{Krivov1998, Kobayashi2008, Kobayashi2009, vanLieshout2014, Sezestre2019}. However, this model fails because the migration timescale dwarfs the survival timescale of hot dust; grains spend the majority of their lifetimes slowly migrating inwards (at lower temperatures), before rapidly sublimating (briefly reaching the required high temperatures) then blowing away (quickly cooling). This results in the PR-drag model producing far too much MIR emission relative to NIR to be compatible with observations.

This problem is partially mitigated by an alternative model, where grains are deposited close to the star by star-grazing comets \citep{Bonsor2014, Raymond2014, Marboeuf2016, Faramaz2017, Sezestre2019}. This scenario is commonly suggested in the literature. It produces considerably less MIR emission than the PR-drag scenario, because it bypasses the evolutionary phase where grains slowly spiral inwards at comparatively cool temperatures. However, current cometary models also fail because they produce too much MIR emission, particularly for A-type stars (Figure 12 in \citealt{Sezestre2019}). This is because, despite grains being hot enough to emit strongly in the NIR upon release from the comet, these grains then produce copious MIR emission as they move away from the star and cool; \cite{Pearce2020} showed that even \mbox{${0.2\um}$-radius} grains released close to an A0V star cannot produce sufficient NIR vs. MIR emission due to this effect (their Figure 9). 
 
The difficulties faced by these models led to the hypothesis that some trapping mechanism may also operate in hot-exozodi systems, which holds grains close to the star and protects them from blowout and sublimation. One model that has been reasonably successful is gas trapping, where gas released by sublimating dust traps incoming grains just exterior to the sublimation radius \citep{Pearce2020}. This model reproduces a broad range of observational constraints on hot dust, and can fully explain the phenomenon for Sun-like stars. However, the model in its current form fails for A-type stars (for which trapped grains are ${\sim5}$ times too large to reproduce observations), and it is unclear whether the model can reproduce the variability demonstrated by at least one hot exozodi (${\kappa \; \rm Tuc}$; \citealt{Ertel2014, Ertel2016}). An alternative mechanism of magnetic trapping has also been proposed, where charged grains are held by stellar magnetic fields \citep{Czechowski2010, Su2013, Rieke2016, Stamm2019}. However, magnetic trapping models in their current forms also fail to explain observations because the mechanism may cease to be effective if grain sublimation is included, and the expected correlations between hot-dust detection rate and magnetic field strength or stellar rotation are absent \citep{Kral2017Review, Kimura2020}. Another trapping mechanism has been proposed involving the Differential Doppler Effect (DDE), but has so far proved ineffective \citep{Sezestre2019}.

Regardless of whether trapping occurs, comets seem a promising mechanism for delivering dust to the hot-emission region near stars. Sun-grazing comets exist in the Solar System, and there is evidence for extrasolar equivalents (`Falling Evaporating Bodies' or FEBs; \citealt{Ferlet1987, Beust1990}). A tentative relation between hot\change{-}dust detections and circumstellar gas indicative of cometary activity has also been suggested \citep{Rebollido2020}, and stochastic comet infall rates would naturally explain the NIR variability seen in at least one system \citep{Ertel2014, Ertel2016}. There are \change{several} known mechanisms capable of producing star-grazing comets including direct injection of Oort-cloud-like comets \citep{Fernandez2021}, inward scattering of material by chains of planets \citep{Bonsor2014}, resonant driving of debris eccentricities by moderately eccentric planets \citep{Faramaz2017}, and secular driving of debris eccentricities by highly eccentric perturbers \citep{Pearce2021}. \change{Comets also appear to be the dominant source of warm zodiacal dust in the Solar System (\citealt{Rigley2022} and refs. therein).} 

The main problem with cometary supply (without trapping) as a hot-exozodi explanation is that grains released near the star would emit too much MIR as they escaped and cooled. However, \cite{Sezestre2019} and \cite{Pearce2020} showed that cometary dust could yield sufficiently high NIR/MIR flux ratios if the dust outflow were somehow truncated, so that only grains close to the star contributed significant emission. Such truncation could occur through grain sublimation. Figure 9 in \cite{Pearce2020} shows that grains with radii ${\leq0.1\um}$ released close to an A0V star would produce NIR/MIR flux ratios consistent with hot\change{-}dust observations, because these grains would fully sublimate before they could escape and cool. This effect was not significant in the investigation of cometary supply by \cite{Sezestre2019}, because \change{their ejecta had a fixed} size-distribution proportional to ${s_0^{-3.5}}$ (where ${s_0}$ is the grain radius upon ejection), with ${s_0}$ \change{ranging from} ${1\nm}$ to ${1\mm}$; whilst some of their grains would sublimate before they could escape and cool, such grains were too short lived and too few in number to dominate emission. A steeper ejecta size-distribution would increase the number of small grains and thus the NIR emission, potentially allowing the comet\change{ary} delivery scenario to work without trapping.

In this paper we revisit and extend the cometary supply model, to determine the comet and ejecta properties required to reproduce hot-dust observations without trapping. If these proved physically viable then comets could be solely responsible for hot exozodis; if not, then some trapping mechanism \textit{must} also operate if hot dust is supplied by comets. The modelling complements previous analyses by \cite{Sezestre2019} by examining broader ranges of comet parameters and dust size-distributions. In particular, we test ejecta size-distribution slopes steeper than \change{that of \cite{Sezestre2019}} (to investigate the effect of sublimation before escape), and consider a broad range of comet eccentricities and pericentres\change{.} However, we show that dust released by star-grazing comets would need to be deposited at unreasonably high rates and with unphysically steep size distributions to replicate hot-exozodi observations, and conclude that cometary supply (without trapping) is unlikely to be responsible for the phenomenon.

The paper layout is as follows. Section \ref{sec: hypothesisAndObservations} describes our cometary supply hypothesis. Section \ref{sec: simulations} details our simulations, and the dust and comet properties required if the mechanism is to operate. These requirements are discussed with regard to the viability of the mechanism in Section \ref{sec: discussion}, along with the model validity and potential future extensions, and we conclude in Section \ref{sec: conclusions}.

\section{Observational constraints and the cometary supply hypothesis}
\label{sec: hypothesisAndObservations}

\subsection{Observational constraints}
\label{subsec: obsConstraints}

We provide a detailed summary of hot-exozodi observational constraints in Appendix \ref{app: obsConstraints}, but here we summarise the constraints that our models are tested against. The spectral energy distributions (SEDs) of hot exozodis are obtained through interferometry and generally have NIR emission at ${\sim1\percent}$ of the stellar level, with significantly lower absolute MIR and FIR emission (e.g. \citealt{Mennesson2014}). Specific fluxes and observation wavelengths differ between systems, but hot exozodis can typically be characterised as having NIR emission in the \change{\textit{H}} and \change{\textit{K}} bands (centred on 1.7 and ${2.2\um}$ respectively) at least \mbox{10 times} higher than MIR emission in the \change{\textit{N}} band (specifically measurements at 8 to ${13\um}$; \citealt{MillanGabet2011, Mennesson2014}). Whilst there may be degeneracies in the MIR \change{\textit{N}}-band interferometric fluxes due to inner-working-angle (IWA) effects at small separations, recent observations of the ${\kappa \; \rm Tuc}$ excess in the MIR \change{\textit{L}} band\change{\footnote{\change{${\kappa \; \rm Tuc}$ is the only hot exozodi with VLTI/MATISSE data.}}} also show a clear flux decline with wavelength between 3.37 and ${3.85\um}$, indicative of an excess SED peaking in the NIR \citep{Kirchschlager2020}. These MIR \change{\textit{L}}-band data have an IWA comparable to the NIR data, with a spectral slope consistent with the excess flux at ${2.2\um}$ being ${\gtrsim 10}$ times that at ${8.5\um}$ (see Section \ref{subsec: discussionInnerWorkingAngle}). We therefore judge the success of any hot-exozodi model on whether it can achieve \change{spectral} flux densities (${F_\nu}$) at ${2.2\um}$ of at least ${1\percent}$ of the stellar level at that wavelength, and whether this ${2.2\um}$ excess flux exceeds \change{that} at ${8.5\um}$ by an order of magnitude (i.e. ${F_\nu(2.2\um) / F_\nu(8.5\um) \gtrsim 10}$). 

\subsection{Cometary supply hypothesis}
\label{subsec: cometHypothesis}

We test the hypothesis that hot dust is cometary ejecta, continually released by star-grazing comets passing pericentre as they undergo processes such as outgassing, sublimation and tidal fragmentation. We do not consider any additional dust-trapping mechanisms. We consider both A0V and G2V stars, because NIR-excesses are common irrespective of spectral type \citep{Ertel2014}. The comets are assumed to be Sun-grazer analogues, with high eccentricities and small pericentres; this class of Solar-System comets can have pericentres less than ${0.01\au}$ (${2\rSun}$) and eccentricities ${>0.9999}$ (e.g. \citealt{Kreutz1888, Opik1966, Marsden1967, Marsden1989, Marsden2005}), and tend to peak in brightness around ${0.06\au}$ (${12\rSun}$; \citealt{Biesecker2002}). Such small pericentres would allow released dust to reach temperatures sufficient for strong NIR emission.

We assume the most optimistic dust composition with the best chance of reproducing hot-exozodi observations: carbon spheres with zero porosity. Carbonaceous materials appear best able to reproduce hot-exozodi SEDs, whilst silicates produce too much MIR emission due to strong spectral features at ${\sim10\um}$ \citep{Absil2006, Akeson2009, Kirchschlager2017}. Assuming non-porous spheres maximises the time grains can remain hot and close enough to the star to emit in the NIR, because different assumptions would increase grain blowout speeds and reduce grain temperatures \citep{Kirchschlager2013, Brunngraber2017}. We leave the ejecta size-distribution as a free parameter. The model assumptions are discussed in Section \ref{subsec: discussionModelAssumptions}.

\section{Dynamical simulations and SEDs}
\label{sec: simulations}

We explore the cometary delivery scenario by running a suite of dynamical simulations to produce simulated SEDs, which we then compare to observations. We use this analysis to determine the system and dust criteria that must be satisfied for cometary supply to explain hot exozodis without the need for trapping. This section only identifies these necessary criteria; the plausibility of these requirements and the resulting viability of the scenario will be discussed in Section \ref{subsec: discussionModelViability}. Section \ref{subsec: simSetup} describes the numerical processes, and Section \ref{subsec: resultsRequirements} describes the comet, ejecta and system properties that must be satisfied for this mechanism to reproduce hot-exozodi observations.

\subsection{Numerical setup and implementation}
\label{subsec: simSetup}

We simulate dust released from a star-grazing comet at pericentre, following the trajectories of grains as they evolve via dynamical forces and the size evolution as grains shrink through sublimation. These dynamical simulations are used to create surface-mass-density maps for different grain sizes, from which we produce simulated SEDs for comparison with observations.

\subsubsection{Dynamical simulations}
\label{subsec: dynamicalSimulations}

Each dynamical simulation comprises a single dust grain and a star, where the grain is initialised with a position and velocity corresponding to the pericentre of a highly eccentric orbit. This initial orbit is assumed to be that of the star-grazing comet from which the grain is released. The grain's subsequent evolution is simulated using a bespoke dynamical integrator, which we created to investigate the gas-trap mechanism for hot dust in \cite{Pearce2020} (we omit any dust-gas interactions here; see Section \ref{subsec: discussionBrakingViaGasOrBFields}). The simulated grain evolves under gravity, radiation forces and sublimation (we also ignore stellar winds; see Section \ref{subsec: discussionWindsAndBFields}). We refer the reader to \cite{Pearce2020} for a detailed description of the integrator, and describe only the included physical effects here.

In addition to the gravitational force $\change{\bm{F}}_{\rm grav}$, the grain experiences a radiation force

\begin{equation}
\change{\bm F}_{\rm rad} = \beta |\change{\bm F}_{\rm grav}| \left[\left(1 - \frac{\dot{r}_{\rm d}}{c} \right)\change{\bm \hat{r}} - \frac{\change{\bm v}_{\rm d}}{c}\right],
\label{eq: radiationForce}
\end{equation}

\noindent where $\beta$ is the ratio of radiation pressure to the gravitational force, $\change{\bm v}_{\rm d}$ is the dust velocity, $\dot{r}_{\rm d}$ is the radial component of $\change{\bm v}_{\rm d}$, $\change{\bm \hat{r}}$ is the radial unit vector and $c$ is the speed of light \citep{Burns1979}. In our simulations this force predominantly manifests itself as radiation pressure, a force directed radially outwards that counteracts the force of gravity. The strength of this force depends on $\beta$, which varies with grain size, composition and morphology. Smaller grains typically have larger $\beta$ values and are thus more affected by radiation forces.

We model grain sublimation using the prescription of \citet{Lebreton2013}, which is based on that of \citet{Lamy1974}. This gives a sublimation rate of

\begin{equation}
\frac{{\rm d}s}{{\rm d}t} = - \gamma \sqrt{\frac{k_{\rm B} T_{\rm d}}{2 \upi \mu m_{\rm u}}} \frac{\rho_{\rm eq}}{\rho_{\rm d}},
\label{eq: sublimationRate}
\end{equation}

\noindent where $s$, $t$, $k_{\rm B}$, $T_{\rm d}$, $\mu$, $m_{\rm u}$ and ${\rho_{\rm d}}$ denote grain radius, time, the Boltzmann constant, dust temperature, the molecular weight of sublimating material, the atomic mass unit and dust density respectively. The value $\gamma$ is an empirical correction factor (we use $\gamma = 0.7$ as in \citealt{Lamy1974}), and $\rho_{\rm eq}$ is the density of sublimated gas at saturation pressure:

\begin{equation}
\log_{10}\left( \frac{\rho_{\rm eq}}{\rm g \: cm^{-3}}\right)= B - A\left( \frac{T_{\rm d}}{\rm K}\right)^{-1} - \log_{10}\left(\frac{T_{\rm d}}{\rm K}\right),
\label{eq: gasDensityAtSaturationPressure}
\end{equation}

\noindent where $A$ and $B$ are material-specific quantities determined empirically \citep{Lebreton2013}. We assume grains are some form of carbon that sublimates according to ${\rm C_1}$-atom emission as in the sublimation of graphite, like \cite{Lebreton2013} and \cite{Sezestre2019}\footnote{Graphite also sublimates via emission of carbon molecules and clusters (C$_n$, where $n > 1$; \citealt{Zavitsanos1973}). However, we neglect these additional channels (as well as other processes such as sputtering) for the reasons discussed in Section \ref{subsec: discussionAdditionalMassLossProcesses}.}. For this assumption the relevant coefficients are $A = 37215$ and $B = 7.2294$ \citep{Zavitsanos1973}. We refer the reader to \cite{Pearce2020} for a derivation of Equation \ref{eq: sublimationRate}.

Each simulation is initialised with a star and a single dust grain, with the above dynamical and sublimation physics implemented. The grain is released close to the star, with a velocity equal to that at pericentre of a high-eccentricity, ${\beta = 0}$ orbit. This corresponds to a grain released at pericentre from a star-grazing comet. Since dust has ${\beta > 0}$, the grain is either unbound from the star upon release or (if bound) has a higher eccentricity than the comet. A grain is unbound if ${v_{\rm d}^2 \geq 2 G M_* (1-\beta)/r_{\rm d}}$ (where $r_{\rm d}$, $G$ and $M_*$ denote grain stellocentric distance, gravitational constant and star mass respectively), so a grain released at comet pericentre is instantly unbound if 

\begin{equation}
\beta \geq \frac{1-e_{\rm c}}{2},
\label{eq: betaForUnbound}
\end{equation}

\noindent where $e_{\rm c}$ is the comet eccentricity. Following release, each simulation is run until the grain either fully sublimates or reaches ${1000\au}$.

\begin{figure}
  \centering
   \includegraphics[width=8cm]{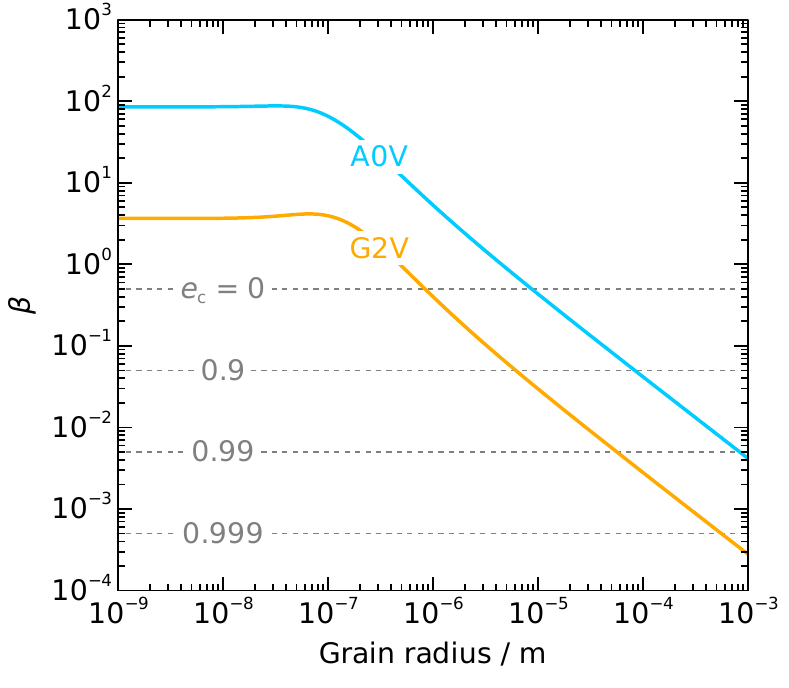}
   \caption{Radiation-pressure coefficient $\beta$ for spherical, solid-carbon grains near an A0V or G2V star (solid lines). A grain released by a comet at pericentre is instantly unbound from the star if its $\beta$ value is above some critical value; dashed lines show these critical $\beta$ values for different values of comet eccentricity $e_{\rm c}$ (Equation \ref{eq: betaForUnbound}).}
   \label{fig: carbonGrainBetasByRadius}
\end{figure}

We take stellar properties from Eric Mamajek's table\footnote{\url{http://www.pas.rochester.edu/~emamajek/EEM_dwarf_UBVIJHK_colors_Teff.txt}} \citep{Pecaut2013}, which yields stellar masses, radii and bolometric luminosities of ${2.30 \: {\rm M}_\odot}$, ${2.09 \: {\rm R}_\odot}$ and ${34.7 \: {\rm L}_\odot}$ for A0V stars and ${1.02 \: {\rm M}_\odot}$, ${1.01 \: {\rm R}_\odot}$ and ${1.02 \: {\rm L}_\odot}$ for G2V stars, respectively. Stellar spectra are from \citet{Kurucz1992}. We consider carbon dust of density ${2 \: {\rm g \: cm}^{-3}}$ and molecular weight 12.01, with optical constants derived from laboratory measurements of \change{carbon produced by pyrolysis of cellulose at $1000\degr{\rm C}$ \citep{Jager1998}\footnote{\url{https://www.astro.uni-jena.de/Laboratory/OCDB/carbon.html}}}. Dust temperature is calculated as a function of grain size and \change{stellocentric} distance using Equation 14 in \citet{Gustafson1994}, and $\beta$ found using \mbox{Equation 3} in that paper; for these equations we use absorption and radiation-pressure efficiencies calculated from the optical constants using either Mie theory \citep{Bohren1983}, Rayleigh\change{-}Gans theory or geometric optics in the appropriate limits (see \citealt{Laor1993} or \citealt{Wyatt2002}). The resulting $\beta$ values are shown on Figure \ref{fig: carbonGrainBetasByRadius}; for carbon, all grains smaller than ${1\mm}$ are unbound if released at pericentre by a \change{comet orbiting an A0V star with eccentricity ${\geq0.99}$, or a comet orbiting a G2V star with eccentricity ${\geq0.999}$}.

\begin{figure*}
  \centering
   \includegraphics[width=17cm]{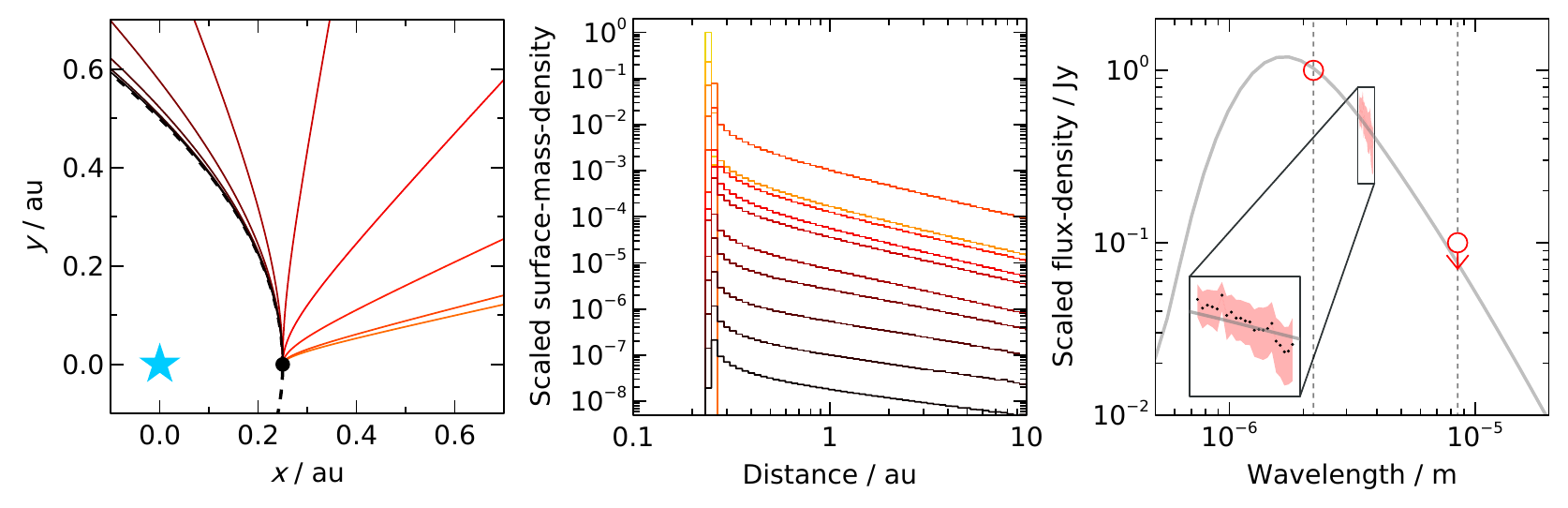}
   \caption{Simulations of grains released at pericentre by star-grazing comets, and the resulting surface densities and SED. The star type is A0V, the comets have pericentre ${0.25\au}$ and eccentricity 0.999, and the grains are solid carbon spheres of initial radii ${10^{-9}}$ (yellow) to ${10^{-3}\m}$ (black), with 13 logarithmically spaced initial grain-sizes shown. Left panel: the dotted line is the comet trajectory, the black circle the dust release point, and the star is at the origin. Solid lines are dust trajectories coloured by initial grain radius. Ejecta with initial radii smaller than ${30\nm}$ released under these conditions sublimate before they are able to travel appreciable distances (Figure \ref{fig: maxSizeForCompleteSublimationVsPeri}) so these trajectories are not visible. Middle panel: azimuthally averaged surface-mass-density profiles in each grain-size bin, assuming new grains are continually released \change{near the pericentre location}. Steps show the distance bin width. One line appears out of colour sequence because that grain-radius bin is populated by grains which start in a larger bin but sublimate down in size. Right panel: SED produced by the entire dust population from the middle panel (stellar flux is omitted), scaled so the ${2.2\um}$ flux is ${1\jy}$. Vertical dotted lines show 2.2 and ${8.5\um}$; to reproduce hot-dust observations, the flux at ${2.2\um}$ (red circle) should be ${\gtrsim 10}$ times that at ${8.5\um}$ (red circle with arrow). Scaled MATISSE data for ${\kappa \; \rm Tuc}$ are also shown around ${3.5\um}$, with pink shading denoting ${1\sigma}$ uncertainties \citep{Kirchschlager2020}; the inset shows these data enlarged. Whilst not part of our formal fitting process, the simulated SED slope is consistent with these data.}
   \label{fig: ejectaTrajectoriesAndSDProfiles}
\end{figure*}

The left panel of Figure \ref{fig: ejectaTrajectoriesAndSDProfiles} shows an example set of simulations run with our integrator. Here comets orbiting an A0V star with pericentre ${0.25\au}$ and eccentricity 0.999 release carbon grains of initial radii ${10^{-9}}$ to ${10^{-3}\m}$ (where each initial grain size is simulated separately, with 19 simulations in total; only 13 simulations are shown for clarity). For this configuration all \change{grains smaller than ${1 \mm}$ in radius} are unbound upon release, with larger grains (${\gtrsim 10 \um}$) following hyperbolic trajectories and smaller grains following anomalous-hyperbolic trajectories (curving away from the star). Only grains with initial radii \change{greater than} ${\change{\sim} 30 \nm}$ escape; smaller grains fully sublimate before they can blow away. This complete sublimation takes 10 to 100 minutes for initial grain radii of 1 and ${10\nm}$ respectively. Escaping grains take 0.01 to ${1\yr}$ to reach ${10\au}$ for initial grain radii of ${0.1\um}$ and ${1\mm}$ respectively.

\subsubsection{Construction of surface-mass-density profiles}

We use these simulations to construct surface-mass-density profiles for different grain sizes; these are azimuthally averaged profiles, produced by assuming a constant dust-input rate at the comet pericentre distance. This corresponds to a scenario where a swarm of comets with identical pericentres and eccentricities, but different orbital orientations and times of pericentre passage, each release dust of various sizes every time they pass pericentre \change{(in reality a single comet would continually release grains, but we assume that it releases the large majority in a narrow a window around pericentre, and that the dynamics of those grains are similar to those released exactly at pericentre)}. We first define sets of grain-size and distance bins. We use 19 grain-size bins with edges spanning ${5.6 \times 10^{-10}}$ to ${1.8 \times 10^{-3}\m}$, spaced logarithmically such that each bin corresponds to one of the simulated initial grain sizes. We use 128 distance bins, spanning from the stellar radius to ${100\au}$. For a given simulation, we calculate how long the grain spends in each size and distance bin as it moves away from the star and sublimates. This yields a set of surface-mass-density profiles for that simulation, each profile corresponding to a different instantaneous grain size (recalling that grain size changes throughout the simulation through sublimation). We then scale the profiles based on the simulation's initial grain size, according to the assumed ejecta size-distribution. Finally, we stack the profiles from all 19 simulations with the same comet parameters. This yields a set of radial surface-mass-density profiles, one for each grain size, that describe the dust released by a family of comets as a function of grain size and stellocentric distance.

The middle panel of Figure \ref{fig: ejectaTrajectoriesAndSDProfiles} shows the surface-mass-density profiles of ejecta with initial radii ${10^{-9}}$ to ${10^{-3}\m}$ released from a swarm of comets orbiting an A0V star with pericentre ${0.25\au}$ and eccentricity 0.999. These profiles are calculated from the simulations on the left panel of that figure (again, only 13 of the 19 profiles are shown for clarity). When scaling each profile we assumed a very steep ejecta size-distribution of ${n(s_0)\change{{\rm d}s} \propto s_0^{-5.5}}\change{{\rm d}s}$, where ${n(s_0)\change{{\rm d}s}}$ is the number of ejecta grains with initial radius \change{in the range $s_0$ to ${s_0+{\rm d}s}$} (the steepness of this size distribution is discussed in Section \ref{subsec: discussionEjectaParameters}). The steep size-distribution means that the smallest grains have the largest \change{surface mass-density}, despite grains initially smaller than ${\sim 30 \nm}$ fully sublimating rather than escaping.

\subsubsection{Calculation of SEDs}

Finally, we use the surface-mass-density profiles to produce simulated SEDs for comparison with observations. The SEDs are generated using \textsc{radmc} \citep{Dullemond2012}; we input the surface-mass-density profiles for all grain sizes, and output a single SED describing the entire ejecta population originating from a single family of comets. We again use stellar spectra from \cite{Kurucz1992}, with dust opacities calculated using the Bohren \& Huffman Mie code\footnote{We test several of our SEDs against equivalents from the more accurate {\sc miex} code \citep{Wolf2004}, and find the two produce similar results in the parameter space of interest.} supplied with \textsc{radmc} (using the optical properties for \change{carbon pyrolysed at $1000\degr{\rm C}$}). We include thermal emission and simple isotropic scattering, although we find that thermal emission dominates at our wavelengths of interest so different scattering prescriptions do not significantly affect the SEDs. We confine dust to a face-on disc with half opening angle ${5^\circ}$ rather than a spherical shell (as could be expected from cometary release) to increase computational efficiency; however, since dust is optically thin and isotropic scattering is used, the simulated SEDs are insensitive to the polar distribution of comet-delivered hot dust. The SED calculation does not account for finite interferometric inner working angles (i.e. it assumes perfect spatial resolution) but this effect is discussed in Section \ref{subsec: discussionInnerWorkingAngle} and Appendix \ref{app: obsConstraints}.

The right panel of Figure \ref{fig: ejectaTrajectoriesAndSDProfiles} shows the SED arising from the dust distributions on the middle panel, with contributions from all dust sizes. Stellar flux is omitted. Thermal emission dominates; the inclusion of scattering only sightly increases the flux at wavelengths below ${1\um}$. This example setup produces an SED with a steep spectral slope and a ${2.2\um}$ flux that is ${\gtrsim10}$ times that at ${8.5\um}$, consistent with hot\change{-}dust observations.

\subsection{Comet, ejecta and system parameters required to reproduce hot-exozodi observations}
\label{subsec: resultsRequirements}

We test the viability of the comet\change{ary} delivery model by performing the above analyses over a range of star, comet and dust parameters, producing a suite of SEDs for comparison with observations. We consider A0V and G2V stars, and test comet pericentres ranging from the stellar radius to ${0.5\au}$ with comet eccentricities up to 0.9999. Ejecta consists of carbon grains with initial size-distribution slopes $q$ ranging from 2.5 to 6.5, where ${n(s_0) \propto s_0^{-q}}$. Ejecta have maximum radii of ${1\mm}$, and we vary the minimum grain radius between ${1\nm}$ and ${0.1\um}$. This section details the comet, ejecta and system properties that must be satisfied for cometary supply to reproduce hot-exozodi observations; we describe the required comet orbits and ejecta size-distributions in Section \ref{subsec: resultsCometOrbitsAndEjectaRequirements} (and explain why they are needed in Section \ref{subsec: resultsUnderstandingCometOrbitsAndEjectaRequirements}), the required comet and dust input rates in Section \ref{subsec: resultsInputRateRequirements}, and the cometary reservoirs in Section \ref{subsec: resultsCometaryReservoirRequirements}. The physical plausibility of these requirements and the resulting viability of the scenario will be discussed in Section \ref{subsec: discussionModelViability}.

\subsubsection{Comet orbits and ejecta size-distributions required}
\label{subsec: resultsCometOrbitsAndEjectaRequirements}

\begin{figure*}
  \centering
   \includegraphics[width=17cm]{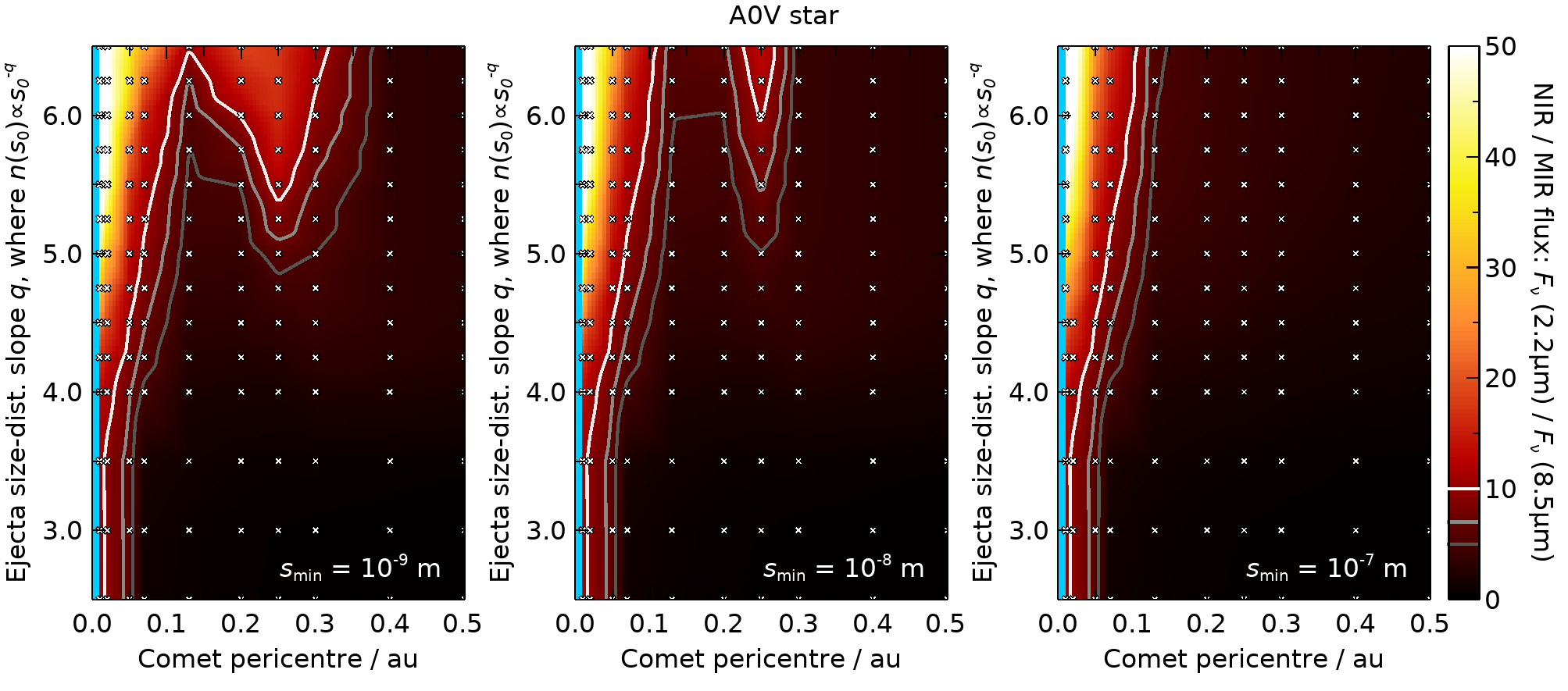}
   \caption{Ratios of NIR/MIR flux from carbon grains released at comet pericentre, for comets with eccentricity 0.9999 orbiting an A0V star. Colours show this ${F_\nu(2.2\um) / F_\nu(8.5\um)}$ flux ratio as a function of comet pericentre and ejecta size-distribution slope, with the lower end of the size distribution truncated at ${1\nm}$, ${10\nm}$ and ${0.1\um}$ for the left, middle and right panels respectively. The largest grains have initial radii ${1\mm}$. The flux ratios are calculated from simulation sets such as that on Figure \ref{fig: ejectaTrajectoriesAndSDProfiles}, where crosses show the tested setups and colours have been interpolated between them. The blue region is the star. To reproduce hot-exozodi observations the dust NIR/MIR flux ratio should be ${\gtrsim10}$ (white line); to achieve this, either comets must pass extremely close to the star (${\lesssim0.02\au}$ from the stellar centre, i.e. ${\lesssim 1}$ stellar radius above the star surface), or the ejecta size-distribution must be very steep upon release and include grains with initial radii ${\lesssim 10\nm}$. In the latter case, comets may have pericentres up to ${0.4\au}$. Physical reasons for the contour shapes are given in Sections \ref{subsec: resultsCometOrbitsAndEjectaRequirements} and \ref{subsec: resultsUnderstandingCometOrbitsAndEjectaRequirements}.}
   \label{fig: a0FluxRatiosForPericentresAndSDSlopes}
\end{figure*}

Observed hot dust produces significantly more NIR than MIR emission, i.e. ${F_\nu(2.2\um) / F_\nu(8.5\um) \gtrsim10}$ (Section \ref{sec: hypothesisAndObservations}). We find that the dust sizes required to produce this flux ratio vary with star type, comet pericentre and eccentricity, but generally the ejecta size-distribution must be very steep. Figure \ref{fig: a0FluxRatiosForPericentresAndSDSlopes} shows our simulated flux ratios as functions of comet pericentre and ejecta  size-distribution slope, for comets with eccentricity 0.9999 orbiting an A0V star. The radii upon release of the smallest ejecta are 1, 10 and ${100\nm}$ on the left to right panels respectively.

The figure shows two regimes where ejecta from these comets yields sufficiently high NIR/MIR flux ratios to reproduce hot-exozodi observations for A0V stars. The first is for small comet pericentres and moderate-to-steep ejecta size-distributions; the required size-distribution slopes range from ${q\approx2.5}$ at ${0.01\au}$ (${1 R_*}$) to the extremely steep 6.5 at ${0.1\au}$ (${10 R_*}$), where $R_*$ is the stellar radius. At these distances many dust sizes are hot enough to produce significant NIR emission, and many grains fully sublimate before they can escape and cool (for comet pericentres less than ${0.06\au}$, all sub-millimetre grains fully sublimate before they can escape). At these small pericentres the flux ratio is insensitive to the smallest ejecta size. 

The second regime yielding the required flux ratio for A0V stars is if comet pericentres are larger (${\sim0.25\au}$; ${26 R_*}$) and the ejecta size-distributions steep (slopes ${q\geq5.5}$), provided that the smallest ejecta has initial radius smaller than ${10\nm}$ (left and middle panels). In this regime the smallest grains dominate the flux, and these sublimate before they can escape and cool (larger grains are too cool to produce sufficient NIR emission at these distances, and they escape before fully sublimating; see Section \ref{subsec: resultsUnderstandingCometOrbitsAndEjectaRequirements}). The reason the size-distribution slope can be shallower at ${0.25\au}$ than at ${0.13\au}$ whilst yielding the same flux ratio is due to the lifetimes of small grains; ${\leq 10\nm}$ grains released at ${0.13\au}$ sublimate ${\sim 7000}$ times faster than those released at ${0.25\au}$, so very steep size distributions are required at the smaller pericentre to ensure enough small grains are present to produce high NIR flux.

Aside from comets with very small pericentres, ejecta is unable to generate sufficient flux ratios for A0V stars if its initial size-distribution slope is shallower than ${s_0^{-5}}$, or if only grains larger than ${10 \nm}$ are released. Furthermore, if comet pericentres are larger than ${0.4\au}$ then no tested size distribution yields sufficient flux ratios, because even ${1\nm}$ grains are too cool to emit sufficient NIR relative to MIR at these distances (see Section \ref{subsec: resultsUnderstandingCometOrbitsAndEjectaRequirements}).

Provided comet eccentricities are very high, the required comet pericentres and ejecta size-distributions do not depend strongly on the exact eccentricity; Figure \ref{fig: a0FluxRatiosForPericentresAndSDSlopes} is largely unchanged if comet eccentricities are reduced from 0.9999 to 0.99 for A0V stars. However, comet eccentricities below this produce significantly lower NIR/MIR flux ratios for all but the smallest comet pericentres or steepest ejecta size-distributions. This is because millimet\change{re}-sized carbon grains are bound upon release for comets with eccentricities below 0.99 around A0V stars (see Figure \ref{fig: carbonGrainBetasByRadius}), so these MIR-producing large grains reside in the system for extraordinarily long times compared to NIR-producing smaller grains (which rapidly sublimate or blow away). For example, a ${1\mm}$ grain released at a pericentre of ${0.13\au}$ by a comet with eccentricity 0.9 orbiting an A0V star survives for ${5000\yr}$ before eventually sublimating and blowing away, compared to a survival time of just ${10^{-8}\yr}$ for a ${10\um}$ grain. Long-lived, large grains therefore dominate emission if comet eccentricities are below ${\sim0.99}$ for A0V stars (except for very steep ejecta size-distributions), so comet eccentricities would have to be higher than this to produce NIR/MIR flux ratios ${\gtrsim10}$.

\begin{figure}
  \centering
   \includegraphics[width=7cm]{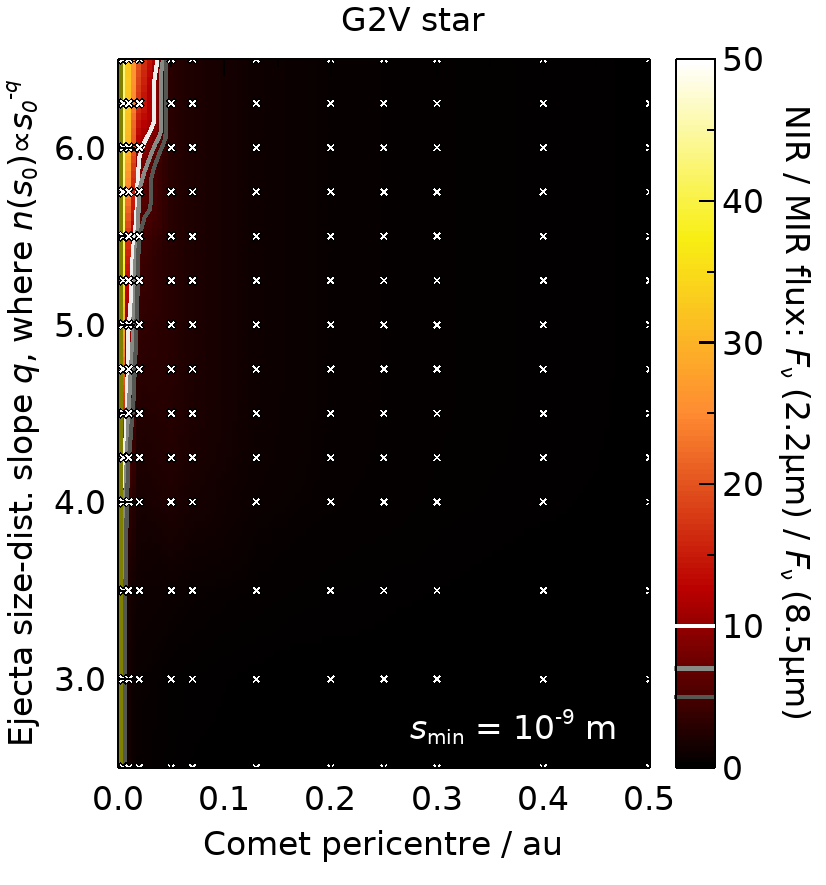}
   \caption{As for Figure \ref{fig: a0FluxRatiosForPericentresAndSDSlopes} (left panel), but for a G2V star. Colours show the ${F_\nu(2.2\um) / F_\nu(8.5\um)}$ flux ratio as a function of comet pericentre and ejecta size-distribution slope, with the lower end of the size distribution truncated at ${1\nm}$. The comet has eccentricity 0.9999, and the stellar radius is ${0.005\au}$ (shaded region). The cometary supply model fails to produce sufficient NIR/MIR flux ratios for G2V stars, unless the ejecta size-distribution is extremely steep (slope ${\geq 4}$) and the comets pass very close to the star (pericentre ${<0.04\au}$, unless the ejecta size-distribution slope is steeper than ${6.5}$).}
   \label{fig: g2FluxRatiosForPericentresAndSDSlopes}
\end{figure}

For G2V stars the hot-exozodi NIR/MIR flux ratio is very difficult to attain. Figure \ref{fig: g2FluxRatiosForPericentresAndSDSlopes} shows this ratio as a function of comet pericentre and ejecta size-distribution slope, for comets with eccentricity 0.9999 orbiting a G2V star. The region of parameter space producing NIR/MIR flux ratios \change{greater than 10} is very small compared to that for A0V stars (Figure \ref{fig: a0FluxRatiosForPericentresAndSDSlopes}); for G2V stars, dust would need to have a very steep size distribution (slope ${>4}$) and be released very close to the star (${\lesssim 0.01\au}$) to produce sufficient flux ratios. The only way to produce a sufficient NIR/MIR flux ratio is to fully sublimate grains before they can escape this region, but this requires dust to pass extremely close to the G2V star. On Figure \ref{fig: g2FluxRatiosForPericentresAndSDSlopes} the lower end of the ejecta size-distribution is truncated at ${1\nm}$, but increasing this minimum size has only a minor effect, equivalent to the slight steepening of the required size distribution for A0V stars at small pericentres (equivalent to pericentres less than ${0.13\au}$ on Figure \ref{fig: a0FluxRatiosForPericentresAndSDSlopes}). G2V-star results do not depend strongly on comet eccentricity for eccentricities larger than 0.9999, but smaller eccentricities produce significantly more MIR emission because the largest, coolest grains are bound upon release and long lived (see Figure \ref{fig: carbonGrainBetasByRadius}). 

To summarise, cometary ejecta must have very steep size distributions upon release to reproduce hot-exozodi observations. Comets must also have pericentres smaller than ${0.4\au}$ (${40 R_*}$) and eccentricities greater than ${0.99}$ for A0V stars. For G2V stars, comets must have pericentres less than ${0.01\au}$ (${2 R_*}$) and eccentricities of at least ${0.9999}$.

\subsubsection{Understanding orbit and size-distribution requirements}
\label{subsec: resultsUnderstandingCometOrbitsAndEjectaRequirements}

\begin{figure*}
  \centering
   \includegraphics[width=18cm]{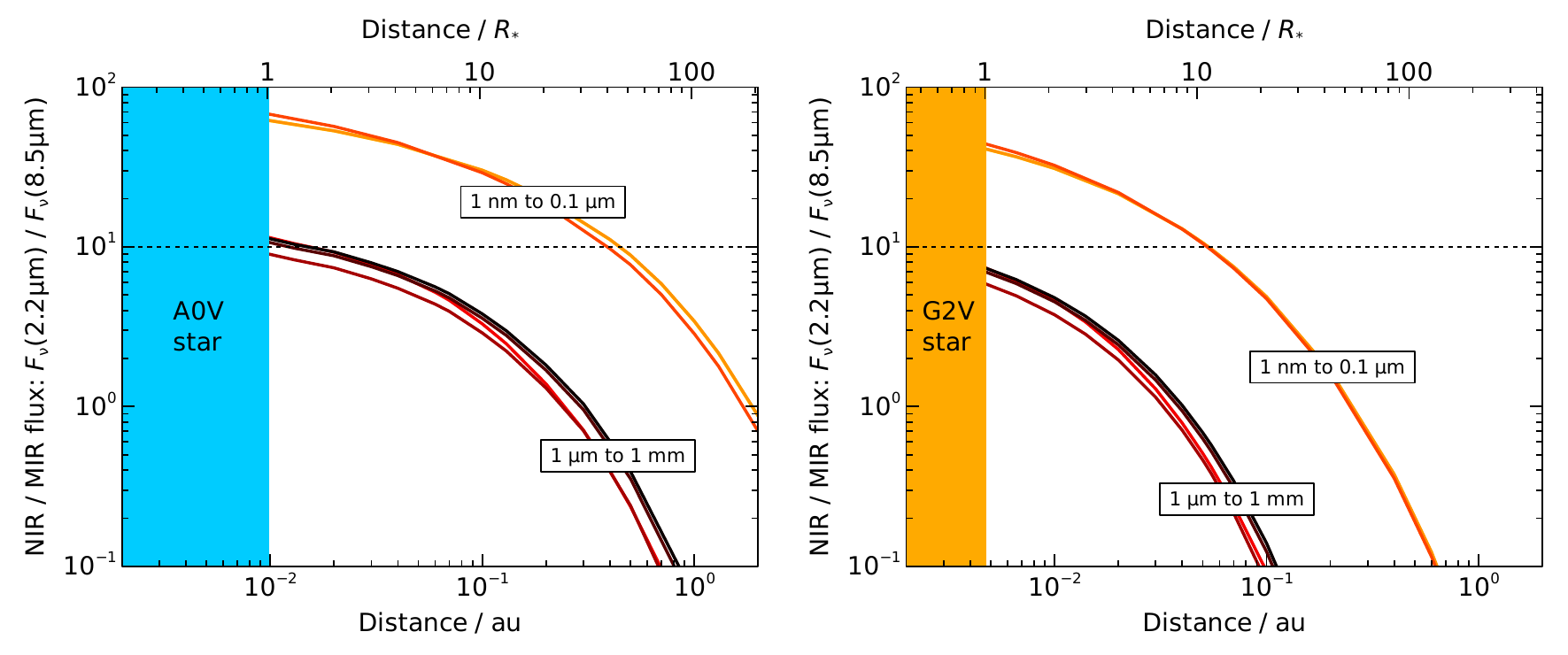}
   \caption{NIR/MIR flux ratio produced by individual carbon grains near an A0V (left) or G2V star (right), as a function of grain size and instantaneous stellocentric distance. The plot is independent of grain dynamics and hot-dust production mechanism. Coloured lines show different grain radii in powers of 10 from ${1\nm}$ to ${1\mm}$; colours are the same as Figure \ref{fig: ejectaTrajectoriesAndSDProfiles}. The flux ratios can be roughly split into two groups: small grains (radii ${1\nm}$ to ${0.1\um}$) and large grains (${1\um}$ to ${1\mm}$), where in each group all grain sizes produce similar flux ratios.  Only sub-micron grains produce flux ratios ${\gtrsim10}$ as required by hot-exozodi observations (dashed lines), unless they are extremely close to an A0V star. No grain produces flux ratios of ${\gtrsim10}$ at distances greater than 0.4 or ${0.05\au}$ from the A0V or G2V star, respectively.}
   \label{fig: fluxRatiosSingleSizeByDistance}
\end{figure*}

The comet orbits and ejecta size-distributions required to reproduce hot-exozodi observations (Section \ref{subsec: resultsCometOrbitsAndEjectaRequirements}) can be better understood by considering the NIR/MIR flux ratios and sublimation behaviours of individual grains. Figure \ref{fig: fluxRatiosSingleSizeByDistance} shows the flux ratio as a function of grain size and instantaneous stellocentric distance, using {\sc radmc} to model emission from a ring of particles of a single size and a Gaussian surface-mass-density profile (full-width-half-maximum of 0.1 times the central radius). This plot is independent of the grain dynamics and hot-dust production mechanism, and depends only on our assumptions about grain composition and optical properties. For A0V stars, the plot shows that only sub-micron grains can produce flux ratios ${>10}$ (aside from larger grains extremely close to the star), and that no grain can produce flux ratios ${>10}$ at distances more than ${0.4\au}$ (${40 R_*}$) from the star. For G2V stars only sub-micron grains produce sufficient flux ratios, and then only when they are interior to just ${0.05\au}$ (${10R_*}$).

\begin{figure}
  \centering
   \includegraphics[width=8cm]{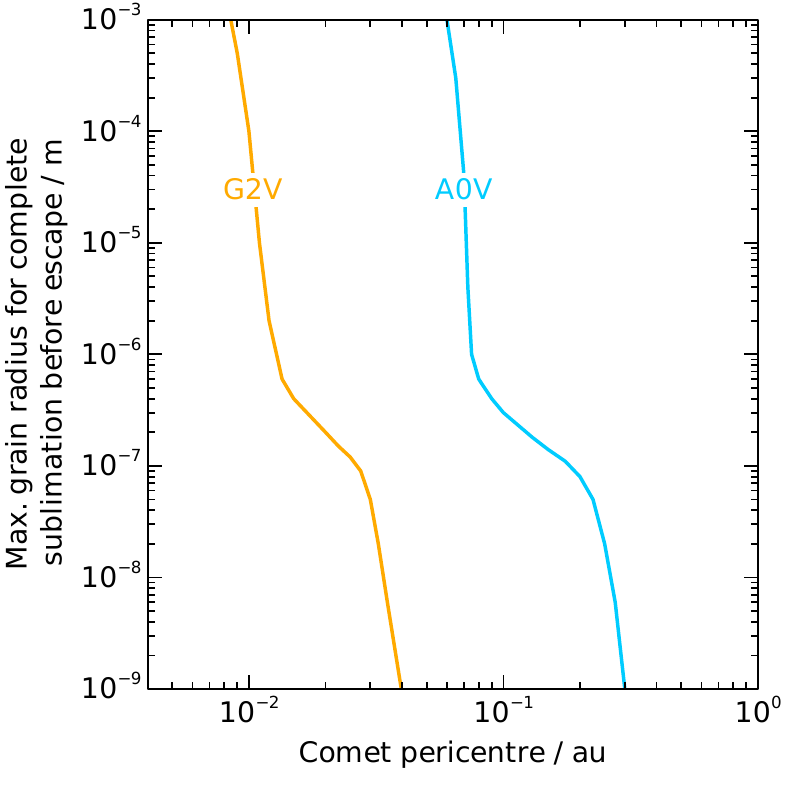}
   \caption{Radii of the largest grains that completely sublimate before escaping, for carbon grains released at pericentre by a comet with eccentricity 0.9999 (Section \ref{subsec: resultsUnderstandingCometOrbitsAndEjectaRequirements}). Blue and yellow lines show A0V and G2V stars respectively.}
   \label{fig: maxSizeForCompleteSublimationVsPeri}
\end{figure}

The sublimation behaviour can be understood from Figure \ref{fig: maxSizeForCompleteSublimationVsPeri}, which shows ejecta sizes that completely sublimate before escaping. The plot is for simulated carbon grains released at pericentre from a comet with eccentricity 0.9999 (these results do not depend strongly on comet eccentricity). For this plot, sublimation before escape is deemed `complete' if the grain size reaches zero before the grain reaches either ${1000\au}$ (for unbound grains) or apocentre (for bound grains). For A0V stars, all sub-millimetre grains sublimate before escaping if released interior to ${0.06\au}$, whilst radii larger than ${0.2\um}$ survive if released outside ${0.13\au}$. All grains larger than ${1\nm}$ survive if released outside ${0.3\au}$. For G2V stars, all sub-millimetre grains fully sublimate if released interior to ${0.008\au}$, and all grains larger than ${1\nm}$ survive if released beyond just ${0.04\au}$. The figure shows that grains released very close to either star sublimate before they can escape and cool, which results in high NIR/MIR flux ratios. For larger release distances, only the smallest grains fully sublimate before escaping; copious quantities of small grains would be needed to produce sufficient NIR emission to overcome MIR from escaping large grains.

\subsubsection{Comet and dust input rates required}
\label{subsec: resultsInputRateRequirements}

\begin{figure*}
  \centering
   \includegraphics[width=17cm]{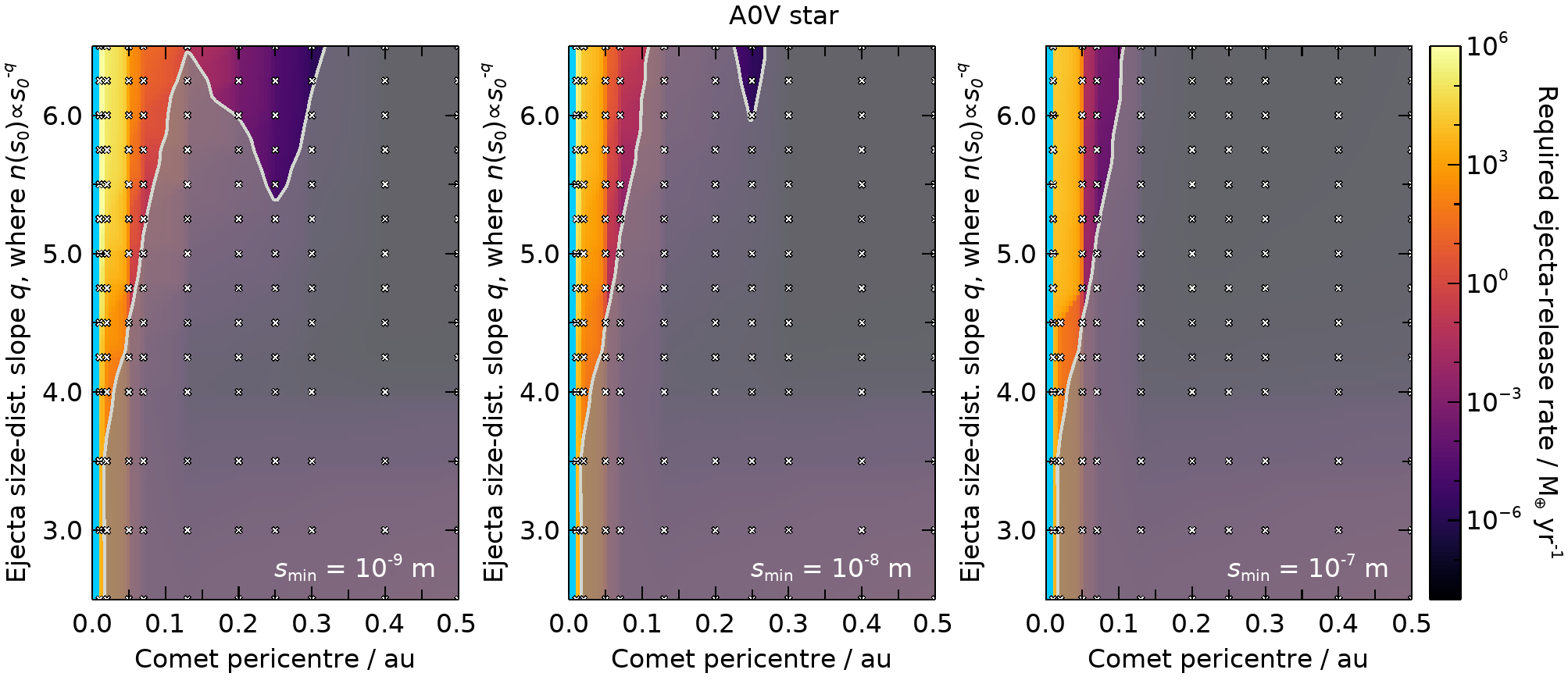}
   \caption{Ejecta-release rates required to sustain a hot\change{-}dust population around an A0V star by comets with eccentricity 0.9999, from the simulations on Figure \ref{fig: a0FluxRatiosForPericentresAndSDSlopes}. These are mass-input rates required to produce a dust flux at ${2.2\um}$ equal to ${1\percent}$ of the stellar flux at that wavelength. The grey shaded region covers setups where dust produces insufficient NIR/MIR flux ratios to reproduce observations (i.e. where ${F_\nu(2.2\um) / F_\nu(8.5\um) < 10}$ from Figure \ref{fig: a0FluxRatiosForPericentresAndSDSlopes}). The plot shows that at least ${10^{-6} \mEarthPerYr}$ of dust must be released close to the star to sustain a hot\change{-}dust population; this equates to \mbox{${3000 \times 10\km}$-radius} comets or \mbox{${20 \times 50\km}$-radius} comets fully disintegrating per year.}
   \label{fig: a0massInflowRatesForPericentresAndSDSlopes}
\end{figure*}

Figure \ref{fig: a0massInflowRatesForPericentresAndSDSlopes} shows the ejecta release rates required to reproduce hot exozodi observations for A0V stars if dust is deposited by comets with eccentricity 0.9999. These are the mass-input rates required such that the simulated dust flux at ${2.2\um}$ equals ${1\percent}$ of the stellar flux at that wavelength, as required by observations (Section \ref{sec: hypothesisAndObservations}).

Total ejecta release rates of at least ${10^{-7}\mEarthPerYr}$ are required to sustain a ${1\percent}$ NIR excess via cometary supply around an A0V star, increasing to at least ${10^{-6}\mEarthPerYr}$ if we omit setups where the NIR/MIR flux ratio is less than 10 (Figure \ref{fig: a0massInflowRatesForPericentresAndSDSlopes}). Assuming a bulk comet density of ${0.5 \gPerCmCubed}$ (as for Solar-System comet 67P/Churyumov-Gerasimenko; \citealt{Patzold2016}), ${10^{-6}\mEarthPerYr}$ equates to \mbox{${3000 \times 10\km}$-radius} comets or \mbox{${20 \times 50\km}$-radius} comets fully disintegrating per year. This rate can only sustain hot-dust populations if comets have pericentres at ${\sim 0.25\au}$ and the smallest ejecta have initial radii of ${10\nm}$ or less, to ensure that the smallest grains produce sufficient emission whilst sublimating slowly enough to produce high NIR/MIR flux ratios. If comet pericentres are smaller than ${\sim 0.25\au}$ then much higher ejecta-release rates are required to counteract rapid grain sublimation (${10^{-3}}$ to ${10 \mEarthPerYr}$ for comet pericentres of ${0.1\au}$, depending on the ejecta size-distribution). These mass-input rates would yield total dust masses of ${10^{-10}}$ to ${10^{-7} \mEarth}$ within ${1\au}$ of an A0V star at any time, consistent with hot-dust masses inferred around A-type stars \citep{Kirchschlager2017}. These masses and input rates were calculated for maximum grain radii of ${1\mm}$, but since the size-distribution slopes are steeper than 4 in the regions of interest, these values are insensitive to maximum grain size. These results are not strongly affected by comet eccentricity.

For G2V stars, only dust released by comets with pericentres less than ${0.04\au}$ can produce sufficient NIR/MIR flux ratios to reproduce observations (unless the ejecta size-distribution is exceedingly steep, i.e. slope greater than ${6.5}$). This arises because significant quantities of dust must sublimate within ${0.05\au}$ of a G2V star to produce sufficient NIR/MIR flux ratios (Section \ref{subsec: resultsCometOrbitsAndEjectaRequirements} and Figure \ref{fig: fluxRatiosSingleSizeByDistance}). This rapid sublimation requires a large dust inflow to sustain a hot exozodi around G2V stars: the mass-inflow rate would have to be greater than ${10^{-4}\mEarthPerYr}$ for hot dust to produce NIR emission at ${1\percent}$ of the G2V-star NIR flux. This corresponds to \mbox{${3 \times 10^5 \times 10\km}$-radius} comets or \mbox{${2000 \times 50\km}$-radius} comets fully disintegrating very close to a G2V star per year. The result would be ${10^{-11}}$ to ${10^{-10}\mEarth}$ of dust within ${1\au}$ at any time, again consistent with hot\change{-}dust masses inferred around Sun-like stars by \cite{Kirchschlager2017}. The result is that the mass-inflow rates required for G2V stars are higher than those for A0V stars in the cometary scenario.

\begin{figure}
  \centering
   \includegraphics[width=8cm]{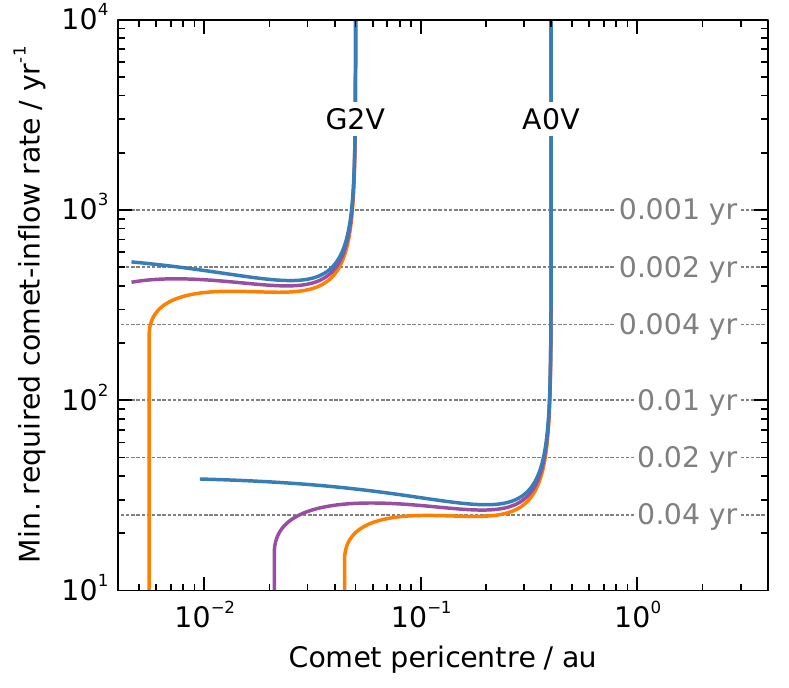}
   \caption{Minimum comet-inflow rate required to sustain a hot\change{-}dust population. This is the minimum rate that comets must arrive to ensure that at least one always lies in the hot-emission region where grains can produce NIR/MIR flux ratios ${\geq10}$; this region is interior to ${0.4\au}$ for an A0V star, or ${0.05\au}$ for a G2V star (Figure \ref{fig: fluxRatiosSingleSizeByDistance}). Solid blue, purple and orange lines show minimum comet-inflow rates for comet eccentricities of 0.9999, 0.9 and 0.8, respectively. Dotted lines show how long a comet spends in this critical region at each pericentre passage. \change{Zero inflow rates are required if the comet apocentre lies inside the hot-emission region (i.e. it would never leave), and lines are truncated if the comet pericentre would lie interior to the stellar radius.}}
   \label{fig: cometInflowRates}
\end{figure}

Aside from the total dust-input rate, a separate constraint is the minimum number of individual comets that must undergo pericentre passage per year to sustain a hot-dust population. For an A0V star, this is the minimum rate that comets must arrive to ensure that at least one always lies within ${0.4\au}$ of the star; since only material within ${0.4\au}$ of an A0V star can produce a NIR/MIR flux ratio ${\geq10}$ (Figure \ref{fig: fluxRatiosSingleSizeByDistance}), and such material rapidly sublimates or escapes, at least one comet must be releasing material within ${0.4\au}$ at any given time to sustain NIR emission. Figure \ref{fig: cometInflowRates} shows this minimum number, \change{found by calculating how much time a comet on an eccentric orbit would spend interior to ${0.4\au}$, then inverting it to get the required rate. The figure suggests} that at least ${30 \cometsPerYr}$ are required to sustain hot-exozodi emission around an A0V star. This is a lower limit; whilst small grains at instantaneous distances of ${0.4\au}$ would produce sufficient NIR/MIR flux ratios, those grains would escape and cool  (Figure \ref{fig: maxSizeForCompleteSublimationVsPeri}), producing too much MIR. Hence grains would have to be released inwards of ${0.4\au}$ to ensure they sublimate before escaping, which increases the required comet-inflow rate; for example, if grains must actually be released interior to ${0.2\au}$, then the required inflow rate increases to at least ${80 \; \rm comets \; yr^{-1}}$ to ensure that at least one comet is always within ${0.2\au}$ of an A0V star. For G2V stars dust must be located within ${0.05\au}$ to produce sufficient NIR/MIR flux ratios (Figure \ref{fig: fluxRatiosSingleSizeByDistance}); since each comet would spend less than just ${1\rm \; day}$ in this region, at least ${400 \cometsPerYr}$ are required to sustain hot-exozodi emission around a G2V star (Figure \ref{fig: cometInflowRates}). Again, this is significantly higher than the required rate for A0V stars, because a comet would spend much longer (${\sim2 \; \rm weeks}$) in the hot-emission region around an A0V star, so fewer comets would be required in that case. These calculations omit the timescales that grains themselves exist in the hot emission region (minutes to hours), because they are much shorter than the time a comet would spend there (weeks).

\subsubsection{Cometary reservoirs required}
\label{subsec: resultsCometaryReservoirRequirements}

There are several ways to estimate the cometary reservoir required to sustain a hot\change{-}dust population. One method is to assume that a hot exozodi is sustained for the entire stellar lifetime. \cite{Kirchschlager2017} constrain hot dust around 11 A-type stars (A0 to A7) with ages ${20\myr}$ to ${1\gyr}$ \citep{Kirchschlager2017, Pearce2022}; since A0V stars require at least ${10^{-6} \mEarthPerYr}$ of carbonaceous dust per year if supplied by comets with eccentricities of 0.9999 (Section \ref{subsec: resultsInputRateRequirements}), this would imply that some A-type stars require cometary reservoirs of at least ${1000\mEarth}$ (even higher in reality, since the comets are unlikely to be ${100\percent}$ carbon). Figure \ref{fig: cometInflowRates} shows that at least ${30\cometsPerYr}$ are required to sustain hot exozodis around A0V stars, implying that some A-type stars would require at least ${3\times 10^{10}}$ individual comets to infall over the stellar lifetime. For Sun-like stars, \cite{Kirchschlager2017} list 8 stars of type F5 to G8 with hot\change{-}dust detections, with ages ${300\myr}$ to ${10\gyr}$ \citep{Kirchschlager2017, Pearce2022}; combined with the minimum ${10^{-4} \mEarthPerYr}$  dust supply required for G2V stars, this yields huge cometary reservoirs of at least ${10^6\mEarth}$ in some cases. Since at least ${400\cometsPerYr}$ would be required (Figure \ref{fig: cometInflowRates}), this would imply that at least ${4\times 10^{12}}$ individual comets fall towards a Sun-like star over its lifetime. These reservoir estimates are lower limits because they assume that all comets fully disintegrate, that all comets in the reservoir eventually become star-grazers, and that comets are ${100\percent}$ carbon without any additional volatiles (which would sublimate at larger distances). The estimates are insensitive to comet eccentricity provided it is large enough for millimet\change{re} grains to be unbound upon release (at least $0.99$ and ${0.9999}$ for A0V and G2V stars respectively); for smaller eccentricities the required reservoirs are larger, because sublimation must be fast enough to quickly remove the otherwise bound grains before they emit too much MIR.

An alternative estimate is that hot dust is only sustained for a fraction of the stellar lifetime. Since NIR excesses are observed for ${\sim20\percent}$ of AFGK-stars and there is no correlation with age (e.g. \citealt{Ertel2014, Kirchschlager2017}), we could hypothesise that a typical star spends ${\sim20\percent}$ of its lifetime with a hot exozodi. Adjusting the above estimates, this implies that some A-type stars require cometary reservoirs of at least ${200\mEarth}$ comprising at least ${6\times 10^{9}}$ individual comets. Likewise some Sun-like stars would require primordial cometary reservoirs of at least ${2\times10^5\mEarth}$ comprising at least ${8\times 10^{11}}$ individual comets.

\section{Discussion}
\label{sec: discussion}

Sections \ref{sec: hypothesisAndObservations} and \ref{sec: simulations} described the observational constraints, cometary supply model and the comet and ejecta parameters required for the model to reproduce hot-exozodi observations. We now discuss the viability of the cometary supply model based on these required parameters (Section \ref{subsec: discussionModelViability}), the assumptions and validity of our modelling approach (Section \ref{subsec: discussionModelAssumptions}), and possible future extensions to the model (Section \ref{subsec: discussionModelExtensions}).

\subsection{Viability of this cometary delivery model}
\label{subsec: discussionModelViability}
 
Section \ref{subsec: resultsRequirements} described the comet, ejecta and system parameters required if dust deposition by star-grazing comets is solely responsible for hot exozodis. High eccentricity comets, small grains, steep ejecta size-distributions, and very large mass inflow rates would be needed. We now consider these requirements to assess the scenario viability; Sections \ref{subsec: discussionCometParametersAndInputRates} and \ref{subsec: discussionEjectaParameters} discuss the comet and ejecta requirements respectively, and Section \ref{subsec: discussionModelViabilitySummary} summarises our conclusions on the model viability.

\subsubsection{Comet orbits, inflow rates and ejecta-release rates}
\label{subsec: discussionCometParametersAndInputRates}

If all ejecta size-distributions were possible, then the comet orbits required to reproduce hot-exozodi observations appear reasonable. Comet pericentres less than ${0.4\au}$ and eccentricities of at least ${0.99}$ are needed for A0V stars, and pericentres less than ${0.01\au}$ and eccentricities of at least ${0.9999}$ are required for G2V stars (Section \ref{subsec: resultsCometOrbitsAndEjectaRequirements}). These are directly comparable with observed Sun-grazing comets, which can have pericentres less than ${0.01\au}$ (${2\rSun}$) and eccentricities above ${0.9999}$ (e.g. \citealt{Kreutz1888, Opik1966, Marsden1967, Marsden1989, Marsden2005}). The required pericentres for A0V stars are also comparable to those inferred for transiting exocomets around the A5V-star ${\beta \; \rm Pic}$, which are typically around ${0.1\au}$ \citep{Kiefer2014, LecavelierDesEtangs2022}\footnote{The star ${\beta \rm \; Pic}$ hosts a hot exozodi \citep{Defrere2012}.}. The required comet orbits therefore appear reasonable.

The requirement that at least ${30\cometsPerYr}$ approach A0V stars for at least one to be in the hot-emission region at any given time (Section \ref{subsec: resultsInputRateRequirements}) also seems reasonable, and the ${400\cometsPerYr}$ required for G2V stars, whilst higher, may also be possible. Both rates are within an order of magnitude of the frequency of Sun-grazing comets, estimated to be ${\sim 90 \; \rm yr^{-1}}$ from \textit{Solar and Heliospheric Observatory} (SOHO) observations \citep{Marsden2005}. A larger rate is inferred for star-grazing exocomets around ${\beta \; \rm Pic}$; at least ${\sim70\cometsPerYr}$  transit that star \citep{Kiefer2014, LecavelierDesEtangs2022}, corresponding to at least 2000 to ${3000\perYr}$ when accounting for non-transiting comets\footnote{We assume comets have an isotropic inclination distribution and eccentricities ${\sim 1}$, then extrapolate the transit rate of ${\gtrsim70\perYr}$ to estimate the non-transiting rate using Equation 9 in \cite{Winn2010} (taking the ${\beta \; \rm Pic}$ radius to be ${1.7\rSun}$; \citealt{Kervella2004}).}.  It is therefore plausible that comet-inflow rates are high enough in hot-dust systems that at least one comet is in the hot-emission region at any one time.

However, the required ejecta-release rates are much less plausible. If all ejecta size-distributions were possible, then dust release rates of at least ${10^{-6} \mEarthPerYr}$ are required to sustain hot-exozodi populations around A0V stars, and at least ${10^{-4} \mEarthPerYr}$ for G2V stars (Section \ref{subsec: resultsInputRateRequirements}). These are the rates required in the region of parameter space where dust produces sufficient NIR/MIR flux ratios to reproduce observations. They equate to the complete disintegration of \mbox{${3000 \times 10\km}$} or \mbox{${20 \times 50\km}$-radius} comets per year for A0V stars, and \mbox{${3 \times10^5 \times 10\km}$} or \mbox{${2000 \times 50\km}$-radius} comets per year for G2V stars. These are much larger than typical Sun-grazing comets; Kreutz Sun-grazers (comprising ${85\percent}$ of SOHO Sun-grazers; \citealt{Battams2017}) have typical radii of just met\change{re}s to tens of met\change{re}s (\citealt{Jones2018}, and refs. therein). Exocomet sizes appear more favourable around ${\beta \; \rm Pic}$; the 30 transiting, star-grazing exocomets inferred by \cite{LecavelierDesEtangs2022} \change{during observations spanning ${156 \; {\rm days}}$} have estimated radii of ${\change{s_{\rm comet}} = 1-10\km}$. Those authors infer a comet size-distribution going as \change{${s_{\rm comet}^{-3.5}}$}, so we could expect transit rates of ${1\perYr}$ for comet radii ${\geq 5\km}$ or ${0.02\perYr}$ for radii ${\geq 25\km}$ for \mbox{${\beta}$ Pic} (corresponding to ${40-50\perYr}$ for ${\geq 5\km}$ or ${0.6-1.0\perYr}$ for ${\geq 25\km}$ after extrapolating to include non-transiting comets). However, these observed rates still fall short of those required by the cometary hot\change{-}dust model, and the latter are \textit{minimum} rates assuming all comets fully sublimate during pericentre passage (and they are ${100\percent}$ carbon). Regardless of how the dust input is considered, it appears that cometary supply is very unlikely to supply enough material to sustain hot exozodis (without additional trapping).

We can also consider the expected gas release during comet sublimation. Our models require at least ${10^{-6}}$ to ${10^{-4} \mEarthPerYr}$ of dust to be released, the large majority of which fully sublimates into gas. This effectively corresponds to a gas-release rate of  ${10^{-6}}$ to ${10^{-4} \mEarthPerYr}$ close to the star, which is much higher than rates currently inferred for exocomet activity. For example, \cite{Beust1989} infer that ${10^{34} \rm \; atoms \; s^{-1}}$ of hydrogen are released per Falling Evaporating Body around ${\beta \; \rm Pic}$, corresponding to ${10^{-10} \mEarthPerYr}$ per comet; to achieve overall gas-release rates of ${10^{-6}}$ to ${10^{-4} \mEarthPerYr}$ as inferred by the cometary hot-exozodi model would therefore require ${10^4}$ to ${10^6}$ such comets per year, which is much higher than the rates inferred from transits. Additionally, if the gas-release rates around all hot-exozodi stars really were higher than that of ${\beta \; \rm Pic}$, then we would expect to detect this gas and see a strong correlation between the presence of hot dust and gas. Whilst a tentative correlation has been suggested, there are many hot-exozodi stars where gas has not been detected \citep{Rebollido2020}. Therefore, the high dust inputs required by the cometary hot-exozodi model could potentially be ruled out by gas considerations alone.

The high dust-release rates also imply that cometary reservoirs of at least 200 to ${10^5\mEarth}$ are needed (Section \ref{subsec: resultsCometaryReservoirRequirements}). Such large reservoirs should be detectable, but many hot-exozodi systems have no detected outer debris populations; indeed, there is no general correlation between the presence of hot dust and FIR excesses associated with cold debris reservoirs \citep{MillanGabet2011, Ertel2014, Mennesson2014, Ertel2018, Ertel2020, Absil2021}. Debris populations more massive than ${100-1000\mEarth}$ may also be physically unfeasible, because these may be larger than the maximum mass of solids that could be inherited from the protoplanetary disc \citep{Krivov2021}. The debris reservoirs required by the cometary supply model therefore appear unfeasibly large, particularly the huge ${10^5\mEarth}$ needed for G2V stars. Our reservoir predictions are also lower limits calculated with very optimistic assumptions about cometary composition and evolution, so the actual reservoirs required by the cometary supply model are probably much larger.

There are additional problems with the overall cometary population. First, unless ejecta size-distributions are very steep, we require eccentricities of at least 0.99 (for A0V stars) or 0.9999 (for G2V stars) to reproduce hot\change{-}dust observations. Whilst such orbits are reasonable for individual comets, we also require the \textit{absence} of comets with smaller eccentricities; it is not sufficient for just some comets to be very eccentric, but rather the majority would have to be. This is because hot dust released by very eccentric comets remains in the hot-emission region for a very short period before sublimating or escaping, so must constantly be replenished. Conversely, grains released by lower-eccentricity comets may be bound upon release, and therefore have much longer lifespans. Such grains would spend most of their time away from the hot-emission region, so would produce copious MIR emission compared to the very short-lived, NIR-producing grains. An example is ejecta with a size-distribution slope of 4 released by comets with pericentre ${0.13\au}$ around an A0V star; the resulting dust flux is 100 times higher if the comets have eccentricity 0.9 rather than 0.9999, and the NIR/MIR flux ratio for the eccentricity 0.9 comets is half that for eccentricity 0.9999. In this case, more than \mbox{${100 \times 0.9999}$-eccentricity} comets would have to pass pericentre for every one 0.9-eccentricity comet passing pericentre in the same period, if the former were to dominate emission. This is unlikely based on Solar-System comets; of the known bound comets with pericentres ${<0.1\au}$, only half of those with eccentricities ${\geq0.9}$ have eccentricities ${\geq0.9999}$ (comet data from the JPL Small-Body Database\footnote{\label{footnote: jplSmallBodyDatabase}\url{https://ssd.jpl.nasa.gov/tools/sbdb_lookup.html\#/}}). Whilst we have not considered observational selection effects in this calculation, it seems unlikely that 100 times more comets with eccentricities ${0.9999}$ pass pericentre for every one with 0.9.

A second issue is that the comet population would decline with time as comets disintegrate, which would imply that hot-exozodi incidence should also decline with stellar age. However, no such correlation is detected \citep{Ertel2014, Kirchschlager2017}. Whilst several mechanisms could increase the production of star-grazing comets later in the stellar lifetime (e.g. \citealt{Faramaz2017} show that planet-debris interactions could produce comets after delays of ${\sim100\myr}$), it seems unlikely that such mechanisms could produce the very high inflow rates required by the comet\change{ary} delivery model.

We also note that the above requirements for comet orbits, inflow rates and ejecta release rates are the most optimistic, valid if all ejecta size-distributions are allowed (including those with very steep slopes of ${q> 5}$ and minimum grain radii of ${10\nm}$ or smaller). However, these ejecta prescriptions may be unrealistic (Section \ref{subsec: discussionEjectaParameters}). If it is only possible to have moderate size-distribution slopes (${\sim 3.5}$) and larger ejecta (${>10\nm}$), then the cometary parameters that can reproduce hot-exozodis become much more constricted. Specifically, cometary supply would not work at all for G2V stars. For A0V stars comets would have to approach within just ${0.02\au}$ (${2 R_*}$), which is closer than pericentres inferred for ${\beta \; \rm Pic}$ comets (typically 0.09 to ${0.18\au}$; \citealt{Kiefer2014, LecavelierDesEtangs2022}) though still possible. However, the mass-inflow rates required in this regime would be vast (${\gtrsim 100 \mEarthPerYr}$; Figure \ref{fig: a0massInflowRatesForPericentresAndSDSlopes}), greatly exceeding theoretical and observational limits.

In summary, whilst the comet orbits required to produce hot exozodis appear reasonable, the dust-input rates are probably far too high to be compatible with current theories on comets and planetary systems. This alone could be sufficient to rule out the model in its current form. The following section shows that the ejecta properties required are also incompatible with our current understanding of cometary dust.

\subsubsection{Ejecta size-distribution and composition}
\label{subsec: discussionEjectaParameters}

Dust from star-grazing comets can only explain hot exozodis if ejecta has a very steep size distribution upon release, typically going as ${n(s_0) \propto s_0^{-5.5}}$ or steeper (shallower distributions are possible, but are probably ruled out by the huge mass-inflow rates required). The smallest ejecta should also have radii in the nanometre range. Such steep size distributions are very different to our understanding of cometary ejecta, being much steeper than the slopes of 3.5 to 4.2 typically assumed (e.g. \change{\citealt{Sekanina1973, Hanner1984, Harker2002}}). There is also evidence that actual size-distribution slopes may \textit{flatten} towards smaller ejecta sizes; Figure 7 in \cite{Blum2017} summarises the ejecta size-distribution for Solar-System comet 67P/Churyumov-Gerasimenko (determined through \textit{Rosetta} and Earth-based observations), suggesting a slope of 3 to 4 around ${10\cm}$ grains but potentially decreasing to just 1 for micron grains. Whilst measurements of nanomet\change{re} grains are lacking, it appears unlikely that the majority of the ejecta mass is contained in nanometre grains for 67P. This comet is not a Sun-grazer (it has pericentre ${1.2\au}$ and eccentricity ${0.64}$; JPL Small-Body Database\textsuperscript{\ref{footnote: jplSmallBodyDatabase}}) so it is possible that a Sun-grazing comet has a steeper ejecta size-distribution around pericentre, although \cite{Kimura2002} show that the tails of Sun-grazers can be reasonably explained with ${0.1\um}$ grains (rather than the nanomet\change{re} grains required for cometary supply alone to reproduce hot exozodis). It therefore appears that the ejecta size-distributions required by the cometary exozodi model are probably unrealistically steep.

Another requirement of current hot-exozodi models is that dust is predominantly carbonaceous; silicates are ruled out because they would produce too much MIR flux from strong emission features around ${10\um}$ \citep{Absil2006, Kirchschlager2017, Sezestre2019}. This means that dust deposited by star-grazing comets should also be predominantly carbonaceous if it is to reproduce observations. However, spectroscopy of Solar-System comets indicate that they contain mixtures of carbon and silicates (e.g. \citealt{Bregman1987, Hanner1994, Hayward2000}), and the silicate component would somehow have to be removed to replicate hot-exozodi observations. Silicates sublimate at lower temperatures than carbonaceous material (${\sim 1000\K}$ rather than ${\sim 2000\K}$; \citealt{Lebreton2013}), so silicates in a star-grazing comet would be expected to sublimate at larger distances than carbon\footnote{For example, various silicates released by Sun-grazing comets appear to sublimate around 7, 11.2 and ${12.3\rSun}$ \citep{Kimura2002}, compared to carbon-sublimation distances of ${\sim5 \rSun}$ expected from our models (Figure \ref{fig: maxSizeForCompleteSublimationVsPeri}).}, which could mean that only carbon grains survive to enter the hot-emission region very close to the star. One additional possibility is that, since cometary grains appear to be aggregates comprising both carbon and silicates (e.g. \citealt{Greenberg1990}), the silicates could sublimate as the star-grazing comet approaches the star, releasing small carbonaceous grains with a steep size distribution. However, the problem with both possibilities is that carbon would only comprise a fraction of the total comet mass, but the high mass-inflow rates required by the hot-exozodi model are for carbon grains only (Section \ref{subsec: resultsInputRateRequirements}); the inclusion of silicates would significantly increase the cometary inflow required, which already seems unrealistically high (Section \ref{subsec: discussionCometParametersAndInputRates}). Whilst it is possible that star-grazing comets in hot-exozodi systems have lower silicate-to-carbon ratios than Solar-System comets, this possibility is difficult to justify.

\subsubsection{Overall model viability}
\label{subsec: discussionModelViabilitySummary}

In summary, it appears unlikely that cometary supply alone can be responsible for hot exozodis. Whilst the existence of star-grazing comets with suitable orbits both in the Solar System and other systems is promising, the cometary inflow rates and ejecta size-distributions required to reproduce observations seem incompatible with our current understanding of comets and planetary systems. It is possible that additional trapping mechanisms operate in conjunction with cometary supply to produce hot-exozodis (trapping could reduce the required dust-inflow rate and ejecta size-distribution slope to reasonable values; see Section \ref{subsec: discussionModelExtensions}), but star-grazing comets alone do not appear sufficient to explain the phenomenon.

\subsection{Model assumptions and validity}
\label{subsec: discussionModelAssumptions}

We now discuss some of our modelling assumptions and their implications for the validity of our study.

\subsubsection{Inner working angle and NIR/MIR flux ratio}
\label{subsec: discussionInnerWorkingAngle}

We assessed the model viability by comparing our simulated dust fluxes to observations. Here we consider whether this direct flux comparison is too simplistic, because the observed fluxes were acquired through interferometry. Interferometric observations may not detect excess emission arising from a region too close to the star; specifically, flux from dust located within the interferometer's inner working angle (IWA) may not be detected. The IWA for an observation at wavelength $\lambda$ with interferometric baseline $B$ is

\begin{equation}
{\rm IWA} = 2.06 \times 10^{5} \frac{\lambda}{X B} \; \rm arcsec,
\label{eq: innerWorkingAngle}
\end{equation}

\noindent where $X$ is a constant (typically ${X \approx 4}$ for hot-exozodi observations; \citealt{Absil2013, Kirchschlager2017, Kirchschlager2020}, though see Appendix \ref{app: obsConstraints}). The specific location where dust becomes undetectable depends on dust spatial distribution, system distance, and observation wavelength and baseline, but for a typical hot-exozodi system distance of ${15\pc}$ (median from Table 1 of \citealt{Kirchschlager2017}) and a NIR baseline of ${34\m}$ (CHARA/FLUOR baseline for ${2.13\um}$ measurements in \citealt{Absil2006, Absil2013}), Equation \ref{eq: innerWorkingAngle} suggests that NIR emission from dust interior to ${0.05\au}$ (${3\mas}$) may not be visible. The equivalent value for the MIR \change{\textit{N}} band is ${0.08\au}$ (${5\mas}$, assuming effective wavelength ${8.5\um}$ and baseline ${85\m}$ for the Keck Interferometer Nuller; \citealt{MillanGabet2011, Mennesson2014}). So at ${\lesssim 0.1\au}$ it may not be appropriate to assess the model on whether the simulated NIR/MIR flux ratio is ${\gtrsim10}$, because some of the NIR and/or MIR flux may be missed by observations. However, the model already struggles at these separations because the mass-inflow rates required to sustain the observed NIR fluxes are unfeasibly large (Figure \ref{fig: a0massInflowRatesForPericentresAndSDSlopes}), so more detailed consideration of inner working angles is unlikely to improve the viability of the cometary supply mechanism.

The observational inference that hot dust emits significantly more \change{\textit{H}}/\change{\textit{K}}-band flux than \change{\textit{N}} band may be unreliable if emission arises from within the \change{\textit{N}}-band IWA, because some \change{\textit{N}}-band flux would be lost. However, additional evidence supports this inference. One hot exozodi has been detected by VLTI/MATISSE in the MIR \change{\textit{L}}-band, around the \mbox{F6IV-V} star ${\kappa \; \rm Tuc}$ \citep{Kirchschlager2020}. These \change{\textit{L}}-band data not only constrain the excess ${3.5\um}$ flux (between the \change{\textit{K}} and \change{\textit{N}} bands), but also the SED slope between 3.37 and ${3.85\um}$. This slope is negative and steep, declining by ${40\percent}$ between 3.37 and ${3.85\um}$ (Figure \ref{fig: ejectaTrajectoriesAndSDProfiles}, right panel; this decrease is unlikely to be caused solely by the ${14\percent}$ difference in IWA expected between these wavelengths from Equation \ref{eq: innerWorkingAngle}, unless the emission region is very narrow). The MATISSE data therefore independently imply that hot-dust emission peaks in the NIR. They are also consistent with the \change{\textit{H}}/\change{\textit{K}}-band flux being significantly higher than the \change{\textit{N}} band; fitting a physically motivated dust emission model to this \change{\textit{L}}-band data and VLTI/PIONIER \change{\textit{H}}-band data (\citealt{Kirchschlager2020}, their model `c') yields an expected 2.2 to ${8.5\um}$ flux ratio of 13, in line with order-of-magnitude estimates for other systems from the \change{\textit{N}} band\footnote{Combining \change{\textit{H}}- and \change{\textit{L}}-band data is complicated for ${\kappa \; \rm Tuc}$ because the \change{\textit{H}}-band excess is variable. We use model `c' of \citet{Kirchschlager2020}, combining \change{\textit{L}}-band and high-flux \change{\textit{H}}-band data, because it appears the most physically plausible. An alternative model combining \change{\textit{L}}-band with low-flux \change{\textit{H}}-band data (their model `d') is also consistent with MATISSE uncertainties, but the \change{\textit{L}}-band spectral slope is systematically steeper than that model. The observed slope is also systematically steeper than (though consistent with) model `c', so the NIR/MIR flux ratio could be even greater than 13.}. The IWA of the MATISSE \change{\textit{L}}-band measurements is ${0.04\au}$ (${2\mas}$; for the longest baseline of ${95\m}$ and distance of ${21\pc}$; \citealt{Gaia2018}), comparable to the VLTI/PIONIER measurements in the \change{\textit{H}}-band for that system \citep{Ertel2014, Ertel2016} but smaller than the IWA expected in the \change{\textit{N}} band (no \change{\textit{N}}-band data are available for ${\kappa \; \rm Tuc}$). Hence \change{\textit{L}}-band MATISSE data place tighter MIR constraints on hot exozodis than \change{\textit{N}}-band data because the former are less affected by the IWA, but these \change{\textit{L}}-band data still support the inference that hot exozodis have ${F_\nu(2.2\um)/F_\nu(8.5\um) \gtrsim 10}$. Our use of this criterion to assess the viability of hot-exozodi production mechanisms therefore seems reasonable.

\subsubsection{Mass-loss processes for hot grains}
\label{subsec: discussionAdditionalMassLossProcesses}

We assume grains only lose mass through sublimation, via emission of ${\rm C_1}$ atoms. This allows direct comparison with the hot-dust studies of \cite{Lebreton2013} and \cite{Sezestre2019}. The prescription is approximate and omits additional channels and mass-loss processes, but is expected to be reasonable for our purposes.

The prescription uses the parametrisation of equilibrium vapour-pressure by \citet{Zavitsanos1973} for graphite between 2440 and ${3000\K}$, which we assume to be valid down to ${\sim 1500\K}$. It models sublimation via atomic-carbon emission (C$_1$), neglecting emission of carbon molecules or clusters (C$_n$, where $n > 1$). This is reasonable for vacuum conditions and temperatures of ${\sim2000\K}$ or lower; while C$_3$ is a major component of carbon vapour at equilibrium \citep{Zavitsanos1973}, C$_3$ emission is less efficient than that of C$_1$ by a factor of five or more \citep[][and refs. therein]{Frolov2022}. Setting the C$_1$ vaporisation coefficient ${\gamma = 0.7}$ (Equation 2) allows direct comparison with \citet{Lebreton2013} and \citet{Sezestre2019}; however, the accuracy of this is uncertain because the vaporisation coefficients for C$_1$ and C$_n$ are not firmly determined. The C$_1$ coefficient may be smaller than 0.7 by a factor of two \citep[][and refs. therein]{Frolov2022}, so we may overestimate mass loss through sublimation. We note, however, that our prescription well-reproduces the empirical sublimation rate of nanometre carbon \citep{Long2020}. Whilst our sublimation prescription may be underestimated and cometary carbon may not be graphitic (e.g. \citealt{Woodward2021}), in the absence of better solutions our assumption that sublimation proceeds via graphite-like C$_1$ emission appears reasonable.

Other processes may also contribute to mass loss from hot grains. These include physical sputtering (removal of atoms through collisions with stellar-wind particles), chemical sputtering (removal of atoms through chemical reactions with incident particles), radiation-enhanced sublimation (RES; a sublimation process induced by the penetration of a particle), and grain-grain collisions. Physical sputtering is less important than sublimation for nanometre-sized carbon grains in the Sun's inner heliosphere, although it can effectively destroy silicates if coronal mass ejections are considered \citep{Baumann2020}. For graphite irradiated by H+ ions (with the flux and energy expected from the solar wind at ${0.05\au}$ from the Sun), erosion by chemical sputtering exceeds that by sublimation at temperatures between 500 and ${900\K}$, and RES dominates between 1000 and ${2000\K}$ \citep{Paulmier2001}. Grain-grain collisions could also erode dust, as discussed in Section \ref{subsec: discussionCollisions}. We omit these additional mass-loss processes in our simulations because they depend on parameters that may be specific to individual stars (for example, stellar-wind speed and density), and also to provide the most favourable setup to test cometary supply of hot dust. The omission of such mechanisms means that our simulated mass-loss rates are probably underestimated.

Including these additional mass-loss channels is unlikely to increase the viability of the cometary hot-dust model. Faster mass loss would allow comets to have slightly larger pericentres, because grains released at larger distances could still sublimate before escaping (as required to produce sufficient NIR/MIR flux ratios). However, this effect would be minor; small grains would still have to sublimate within the hot-emission regions to produce sufficient NIR/MIR flux ratios (within 0.4 or ${0.05\au}$ for A0V or G2V stars respectively; Figure \ref{fig: fluxRatiosSingleSizeByDistance}), but the original sublimation prescription already ensures this behaviour for nanometre grains released only slightly interior to these distances (0.3 and ${0.04\au}$ for A0V and G2V stars respectively; Figure \ref{fig: maxSizeForCompleteSublimationVsPeri}). Therefore, faster mass loss could only slightly increase the allowed comet pericentres. It could also slightly reduce the required ejecta size-distribution slopes, because larger grains could also fully sublimate before they could escape, but this benefit is also expected to be minor because it would be offset by the faster sublimation of smaller grains. The main effect expected from increasing the mass-loss rate is an increase in the dust-inflow rates required to sustain NIR excesses, and these appear implausibly large already (Section \ref{subsec: discussionCometParametersAndInputRates}). In summary, including additional mass-loss channels is unlikely to increase the viability of the cometary supply model, so we expect our conclusions to hold even if additional mass-loss channels operate.

\subsubsection{Stellar winds and magnetic fields}
\label{subsec: discussionWindsAndBFields}

Our models omit stellar winds. Including winds could slightly reduce the minimum required comet eccentricities, because we implicitly require all grains to be unbound to reproduce hot-exozodi NIR/MIR flux ratios; winds would increase the effective $\beta$, so grains would be unbound for lower comet eccentricities than if winds were absent (Figure \ref{fig: carbonGrainBetasByRadius}). However, the magnitude of this effect is unlikely to be large. 

Magnetic fields are also omitted, despite the likelihood that grains become charged and potentially interact with stellar magnetic fields (e.g. \citealt{Rieke2016, Kimura2020}). We omit magnetic fields because this paper tests whether hot exozodis can be produced by cometary supply alone, in the absence of any additional trapping mechanism. In reality magnetic fields would be present and would affect grain dynamics, but it is unclear whether their inclusion would significantly affect our conclusions (Section \ref{subsec: discussionBrakingViaGasOrBFields}). Since hot exozodis are detected across a broad range of spectral types with diverse magnetic field strengths, it remains unclear whether magnetic fields play a significant role in hot-exozodi production \citep{Kimura2020}.
 
\subsubsection{Dust-release point}
\label{subsec: discussionDustReleasePoint}
 
We model dust release only at comet pericentre, whilst in reality dust would be released at all points around the comet orbit (particularly if comets release dust through fragmentation in addition to sublimation; \citealt{Rigley2022}). \change{However, this approximation is unlikely to have affected our conclusions, for two reasons. First, for star-grazing comets the dust-release rate is expected to steeply increase as the comet approaches pericentre; the model of \cite{Marboeuf2016} (as used by \citealt{Sezestre2019}) has release rate scaling with the inverse square of comet distance if close to the star, and an even steeper dropoff further away. Second, dust close to a star is brighter than that further out (in both thermal emission and scattered light), which further weights the overall emission towards grains released at pericentre. Hence our assumption that dust is released exclusively at comet pericentre is probably reasonable.} We check this by comparing our results to \cite{Sezestre2019}, who performed similar modelling but implemented dust release around the entire comet orbit\change{. T}he setup on their Figure 12a (a comet with pericentre ${0.6\au}$ and eccentricity 0.976 orbiting an A0V star, ejecting ${1.7\nm}$ to ${1\mm}$ grains with a size\change{-}distribution slope of 3.5) produces a ${2.2\um}$ to ${8.5\um}$ flux ratio of ${\sim1}$ for carbon grains\change{; this} is similar to our model for a comparable setup\change{, suggesting that our approximation is appropriate. It should also be noted that r}eleasing grains around the entire orbit would be expected to increase MIR emission, and make the model less able to reproduce hot-exozodi observations; including dust release away from pericentre would therefore only strengthen our conclusions that cometary supply alone is unlikely to produce hot exozodis.

\subsubsection{Grain-cooling timescales}
\label{subsec: discussionGrainCoolingTimescales}

We assume dust temperature depends only on grain size and distance, i.e. escaping grains `instantly' change temperature as they move away from the star. However, realistic grains would take time to radiate heat and cool; escaping grains would be hotter than their distance implies, because they would carry residual heat imparted when they were closer to the star. 

\begin{figure}
  \centering
   \includegraphics[width=8cm]{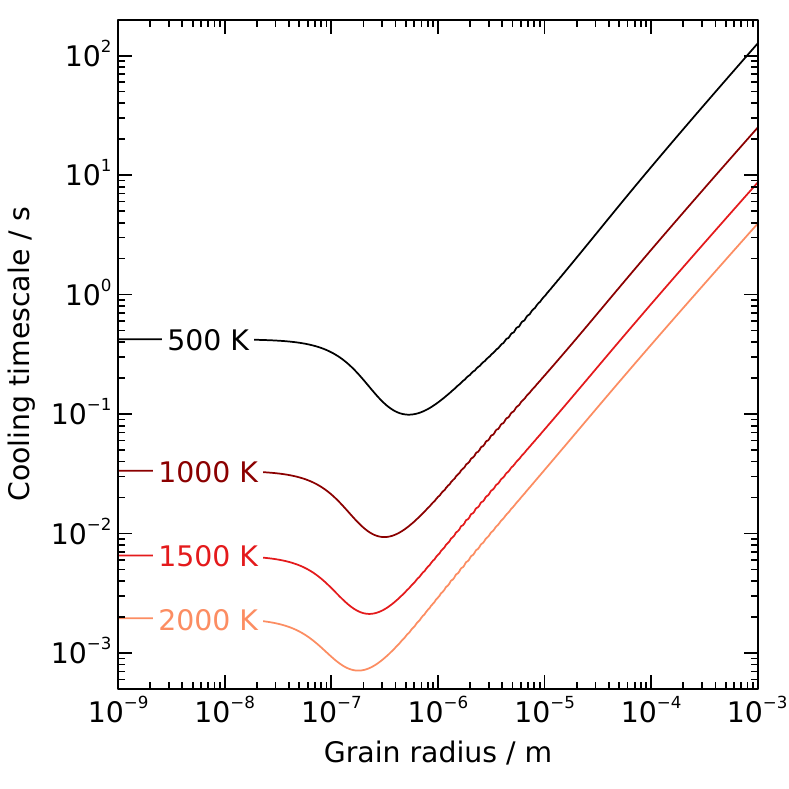}
   \caption{Cooling timescales for graphite grains as functions of grain radius and temperature (Section \ref{subsec: discussionGrainCoolingTimescales}). These are much faster than dynamical timescales, so our assumption that escaping grains `instantly' cool as their distance increases is valid.}
   \label{fig: grainCoolingTimescales}
\end{figure}

We test the importance of this by calculating graphite cooling timescales, i.e. the time it takes internal grain energy to be radiated away following a change in absorption rate (as occurs with changing stellocentric distance). The method is described in Bensberg \& Wolf (submitted). We determine the internal energy using calorimetric data for graphite from \cite{Draine2001}, with emission rates calculated from Mie scattering using {\sc miex} (assuming a 1/3, 2/3 ratio for parallel and perpendicular orientations respectively; \citealt{Draine1993}). We assume all internal energy is radiated without changing temperature. The results are shown on Figure \ref{fig: grainCoolingTimescales}. For all tested grain sizes and temperatures the cooling timescale is ${\lesssim 100 \rm \; s}$, far quicker than timescales for escaping dust to move away from the star (0.01 to ${1\yr}$; Section \ref{subsec: dynamicalSimulations}). This shows that the cooling timescales are much shorter than the dynamical timescales, so our assumption that grains instantly change temperature with distance is unlikely to affect our conclusions.
 
\subsubsection{\change{Omission of planet-dust interactions}}
\label{subsec: discussionOmissionOfPlanets}

\change{We omit planets in our dynamical models, implicitly assuming that ejecta proceed unimpeded through the system. Planets would scatter material that passes within a few Hill radii, but pose little danger to the ejecta we consider. Since only unbound ejecta can reproduce hot-exozodi observations in our scenario, such dust would have near-radial trajectories; therefore, very few grains would encounter planets as they rapidly left the system (especially if their trajectories were also inclined to the planetary plane). The additional requirement that the vast majority of grains sublimate rather than escape means that most ejecta would never even reach planetary distances, further reducing any impact that planets could have. The omission of planets is therefore unlikely to have affected our conclusions.}
\subsubsection{Could we be wrong about NIR and MIR excesses being  contemporary?}
\label{subsec: discussionSimultaneity}

Hot-exozodi models attempt (and struggle) to simultaneously fit NIR and MIR excesses. However, for each target the NIR and MIR excesses are observed at different times by different instruments. There therefore exists the possibility that both NIR and MIR excesses vary simultaneously, in which case NIR and MIR measurements of a single target may correspond to different, incompatible flux states. For example, if we observe NIR when both NIR and MIR are in a high flux state, and later observe MIR when the system is in a low flux state, then we would think that the NIR/MIR flux ratio is much steeper than it actually is. For this to occur hot exozodis would have to vary significantly between observations, which is plausible based on the ${\lesssim 1\yr}$ NIR variability of ${\kappa \; \rm Tuc}$ \citep{Ertel2014, Ertel2016}. This effect could mean that NIR/MIR flux ratios are much shallower than we currently think. However, evidence against this comes from MATISSE observations of ${\kappa \; \rm Tuc}$ \citep{Kirchschlager2020}, that show a clear flux decline around ${3.5\um}$ consistent with a steep spectral slope (Figure \ref{fig: ejectaTrajectoriesAndSDProfiles} right). So whilst follow-up NIR and MIR observations of many hot-exozodi targets are urgently required to better establish variability, it seems unlikely that non-simultaneous NIR and MIR observations have caused us to overestimate the steepness of hot-exozodi spectral slopes.

\subsection{Possible model extensions}
\label{subsec: discussionModelExtensions}

We have shown that the cometary supply model in its current form is unlikely to reproduce hot-exozodi observations. Here we briefly discuss several possible extensions to the model, which could be investigated as ways to explain the phenomenon. These extensions are beyond the scope of this paper, but could be implemented in future works.

\subsubsection{Braking of grains via cometary gas or magnetic fields}
\label{subsec: discussionBrakingViaGasOrBFields}

Sublimating comets and dust grains would release significant quantities of gas. In \cite{Pearce2020} we showed that such gas could trap grains and explain many observed features of hot exozodis, although the model still struggled for A0V stars due to our inclusion of gas accretion onto grains (ultimately leading to the smallest trapped grains being ${\sim 5}$ times too large to reproduce A0V-star observations). However, we also showed that unbound grains could be significantly slowed as they travelled through gas, before ultimately escaping (Figure 7 in \citealt{Pearce2020}). Gas released by sublimating cometary dust could have a similar effect, slowing escaping grains and increasing their time spent in the hot-emission region. This would increase the amount of time grains have to fully sublimate (in the absence of trapping, we require small grains to sublimate before they can escape if they are to produce sufficient NIR/MIR flux ratios), which could allow the cometary supply model to work at slightly larger stellar distances (Figure \ref{fig: maxSizeForCompleteSublimationVsPeri}). In turn, this could reduce the mass-inflow rates required by the cometary model. \change{Magnetic fields could induce a similar braking effect; both gas- and magnetic-braking should be investigated in the star-grazing comet scenario as ways to slow escaping grains, and therefore reduce the mass-inflow rates required to reproduce hot exozodis.}

\subsubsection{Collisions}
\label{subsec: discussionCollisions}

Grain-grain collisions were not included in our models, but are expected to have two effects. Firstly, collisions between unbound, escaping grains could occur when grains are released from comets; such collisions would be expected to be most frequent around comet pericentre, where the population of newly released grains is densest. These collisions could be destructive, particularly if they occurred between grains of different ${\beta}$ values (whose trajectories and velocities could differ considerably). This could disintegrate larger grains just after release, producing smaller grains and thus steepening the effective size distribution close to the star, potentially allowing the initial ejecta size-distributions to be shallower. If these newly produced small grains sublimated before escaping, then the MIR emission would be reduced considerably, potentially making it easier to reproduce hot-exozodi observations.

Secondly, collisions could reduce the minimum comet eccentricities required to produce sufficient NIR/MIR flux ratios. In the non-collisional model, lower comet eccentricities are disfavoured because larger, cooler grains are bound upon release, and therefore have orders-of-magnitude longer lifetimes than the small, NIR-producing grains that rapidly sublimate (Section \ref{subsec: resultsCometOrbitsAndEjectaRequirements}). This results in copious MIR emission, preventing lower-eccentricity comets from reproducing hot-exozodi observations. The lifetimes of such bound, high-eccentricity grains can be thousands of years in our collisionless model, because they spend very little time around pericentre (where they undergo just a brief period of sublimation before moving away from the star). After many orbits, these grains eventually shrink enough that they either fully sublimate at pericentre or blow out of the system. However, such high-eccentricity grains could be expected to undergo violent collisions, which could potentially limit their lifespan and release smaller grains. If the collisional timescales of bound grains were shorter than their sublimation timescales, then including collisions would reduce MIR emission from such grains and potentially allow lower comet eccentricities to also produce sufficient NIR/MIR flux ratios.

\section{Conclusions}
\label{sec: conclusions}

We investigate the supply of hot dust via star-grazing comets (without trapping) as a potential hot-exozodi production mechanism. We simulate the trajectories and size evolution of dust grains released by star-grazing comets at pericentre, for a range of comet orbits, star types and ejecta size-distributions. We find that cometary supply alone is unlikely to reproduce hot-exozodi observations, unless the ejecta properties and cometary populations are very different to our current understanding of comets and planetary systems. In particular, cometary supply without trapping only works as an explanation if the ejecta size-distribution is extremely steep and the dust-inflow rate very high \change{(or unless small grains have very different dynamic and emission properties to larger grains)}. Since continual dust supply through PR-drag is also unable to reproduce hot-exozodi observations, we conclude that simply getting dust close to the star \change{may not be sufficient} to generate hot exozodis, \change{implying} that some mechanism \change{may trap (or at least slow)} hot dust in the vicinity of the star.

\section*{Acknowledgements}

We thank Harald Mutschke, J\"{u}rgen Blum and Cornelia J\"{a}ger for useful discussions\change{, and the anonymous referee whose comments and suggestions improved the paper}. TDP, AVK and MB are supported by the Deutsche Forschungsgemeinschaft (DFG) grants \mbox{Kr 2164/13-2}, \mbox{Kr 2164/14-2}, and \mbox{Kr 2164/15-2}. FK has received funding from the European Research Council (ERC) under the EU's Horizon 2020 research and innovation programme DustOrigin (\mbox{ERC-2019-StG-851622}). GR is supported by DFG project 451244650. SE is supported by the National Aeronautics and Space Administration through the Exoplanet Research Program (Grant No. 80NSSC21K0394). SW is supported by DFG grant \mbox{WO 857/15-2}. \change{JCA is supported by the Programme National de Plan\'{e}tologie (PNP) of CNRS/INSU and the Centre National d'\'{E}tudes Spatiales (CNES), France.}

\section*{Data availability}
\noindent The data underlying this article will be shared upon reasonable request to the corresponding author.


\bibliographystyle{mn2e}
\bibliography{bib_cometaryDelivery}


\appendix

\section{Observational constraints on hot exozodis}
\label{app: obsConstraints}

Detecting exozodiacal dust requires infrared interferometry \citep{Absil2006, Ertel2014, MillanGabet2011, Ertel2018} except in extreme cases of unusually high dust masses \citep[e.g.,][]{Chen2006,Defrere2015}. Optical long-baseline interferometry is used up to \mbox{\change{\textit{L}} band}, while nulling interferometry is used at longer wavelengths. In both cases dust is spatially resolved from the star and, for nulling interferometry, starlight is suppressed to allow measurement of dust brightness. This brightness is commonly expressed as a relative value (a dust-to-star flux ratio or null depth), which is converted into an absolute brightness using certain assumptions \citep{diFolco2007,Kennedy2015}.

Observations have limited inner-working-angle (IWA) and field-of-view (FoV) and, for nulling interferometry, a transmission pattern that must be considered when comparing measurements from different instruments and/or wavelengths. This has implications for the dust locations that can be probed by specific observations, and what fraction of dust emission can be detected at the observing wavelength.

Two instruments are responsible for the vast majority of NIR hot-dust detections: CHARA/FLUOR \citep{Absil2013} and VLTI/PIONIER \citep{Ertel2014,Absil2021}. \citet{Absil2021} show that the NIR IWA is usually small enough to not significantly affect sensitivity for face-on, disc-like dust distributions. For inclined discs or spherical configurations some emission  would be lost inside the IWA. Both effects would result in an under-estimation of dust emission; this would not affect our conclusions, because explaining even detected dust levels is challenging. Dust is also assumed to be fully contained in the interferometric FoV of the NIR observations, which is reasonable since any dust emitting at \change{\textit{H}} or \mbox{\change{\textit{K}} band} must be very hot (hence close to the star). These arguments mean our paper can be agnostic regarding which instrument NIR measurements were obtained with. 

However, for MIR the situation can be very different, since many observations use nulling interferometry. The Large Binocular Telescope Interferometer (LBTI, \citealt{Hinz2016,Ertel2020spie}) was designed for sensitive detections of habitable-zone dust around nearby stars, but its $\sim$40\,mas IWA\footnote{IWA calculated using ${X=4}$ in Equation \ref{eq: innerWorkingAngle} for consistency with other work and this paper, although \cite{Ertel2018, Ertel2020} use a more conservative $X=2$ which would imply an LBTI IWA of ${70\mas}$.} is not sufficient to reliably detect hot dust close to stars. Observations with the Keck Interferometer Nuller (KIN; \citealt{Serabyn2012, Colavita2013}) with an IWA of ${\sim5\mas}$ can better detect hot dust, but at lower sensitivity. The KIN FoV is also limited (${400\mas}$ FWHM; \citealt{Mennesson2014}), but this is unlikely to affect our conclusions since we mostly consider dust emission in both NIR and MIR. For recent VLTI/MATISSE observations in the \mbox{\change{\textit{L}} band} \citep{Kirchschlager2020}, interferometric baselines were chosen to yield a similar IWA as previous FLUOR and PIONIER observations in NIR, so these MATISSE data are compatible with NIR data. For a nulling interferometer an additional complication is the transmission pattern, which consists of stripes of transmissive and dark fringes \citep{MillanGabet2011, Ertel2018spie}; this pattern affects the fraction of dust emission detected, depending on its spatial distribution. However, for well-resolved dust an approximation can be made that ${\sim50\percent}$ of dust emission outside the IWA is transmitted.

NIR excesses must be dominated by hot grains close to the star. However, MIR excesses could either be the Rayleigh-Jeans tail of hot-dust emission, or the emission of cooler, more-distant grains. A lack of MIR emission in KIN and LBTI data for systems with NIR detections is commonly attributed to a lack of  habitable-zone grains, which is considered proof that PR-drag alone cannot explain hot exozodis.

Our paper uses NIR excesses and the lack of MIR emission (from KIN) in those systems to probe the hot-dust production mechanism. In particular we consider the hot-dust SED slope from the NIR to the MIR. For this we must consider systems with relevant data on both, i.e. NIR excesses from FLUOR or PIONIER and MIR data from KIN; there are nine systems with both NIR excesses and KIN observations \citep{Kirchschlager2018}. Of those, only the A3V-type ${\beta \; \rm Leo}$ (${\HD102647}$) has a clear KIN detection \citep{Mennesson2014}.  The star has also been observed with LBTI \citep{Ertel2020} with an IWA of 0.8\,au, yielding an MIR excess consistent with the KIN data. This, together with the LBTI analysis \citep{Defrere2021}, suggests that the \mbox{\change{\textit{N}}-band} MIR excess of ${\beta \; \rm Leo}$ originates from cooler grains relatively far from the star (rather than being the Rayleigh-Jeans tail of hot dust). We thus conclude that no hot dust has so far been detected in \mbox{\change{\textit{N}} band} (around ${10\um}$), and the hot-dust spectral slope steeply declines with increasing wavelength. This is further corroborated by data on ${\kappa \; \rm Tuc}$ \citep{Kirchschlager2020}, which show a steep slope between 3.37 and ${3.85\um}$ (Figure \ref{fig: ejectaTrajectoriesAndSDProfiles} right). A final consideration is that seven of the nine stars from \citet{Kirchschlager2018} are A-type; the other two are $\tau$\,Cet (G8V) and 10\,Tau (F9\,IV-V), so conclusions from this sample are biased toward early-type stars (although the two Sun-like stars do show strong NIR vs. MIR excesses, like the A-types).


\label{lastpage}

\end{document}